\documentclass[12pt]{article} 
\usepackage{graphicx}
\usepackage{soul}
\usepackage{floatrow}
\usepackage{multicol}
\usepackage{latexsym}
\usepackage{float}
\usepackage{stfloats}
\usepackage{subfigure}
\usepackage{array, multirow, bigdelim, makecell, booktabs}
\usepackage[flushleft]{threeparttable}
\usepackage{geometry}
\usepackage{amsmath} 
\usepackage{amssymb} 

\usepackage{authblk}
\usepackage{cite}
\usepackage[numbers,super,sort]{natbib}

\usepackage{notocite}
\usepackage{hyperref}

\def\aj{Astron. J.}
\def\araa{Ann. Rev. Astron. Astrophys.}
\def\aapr{Astron. Astrophys. Rev.} 
\def\apj{Astrophys. J.}
\def\apjl{Astrophys. J. Lett.}
\def\apjs{Astrophys. J. Suppl.}
\def\aap{Astron. Astrophys.}
\def\aaps{Astron. Astrophys. Suppl.}
\def\mnras{Mon. Not. R. Astron. Soc.}
\def\natast{Nature Astronomy}
\def\JApA{J. Astron. Astrophys.}
\def\apss{Astron. Space. Sci.}
\def\pasa{Pub. Astron. Soc. Aus.}

\def\procspie{Proc. SPIE}

\def\aaps{A\&AS}

\title{Extended Far-Ultraviolet Emission in Distant Dwarf Galaxies}

\author[1]{{Anshuman Borgohain} \footnote{Email: ayush.borgohain@gmail.com}}
\author[2]{Kanak Saha \footnote{Email: kanak@iucaa.in}}
\author[3]{Bruce Elmegreen}
\author[1]{Rupjyoti Gogoi}
\author[4]{Francoise Combes}
\author[2]{Shyam N. Tandon}

\affil[1]{Department of Physics, Tezpur University, Napaam 784028, India}
\affil[2]{Inter-University Centre for Astronomy and Astrophysics, PostBag 4, Ganeshkhind, Pune-411007, India}
\affil[3]{IBM Research Division, T. J. Watson Research Center, 1101 Kitchawan Road, Yorktown Heights, NY 10598, USA}
\affil[4]{Observatoire de Paris, LERMA, College de France, CNRS, PSL, Sorbonne University, Paris, France}

\date{}

\begin{document}

\maketitle

{\bf  
Blue Compact Dwarfs (BCDs) are low-luminosity (M$_{K} > -21$ mag)\cite{GildePazetal2003}, metal-poor ($\frac{1}{50}$ $\le Z/Z_{\odot} \le\frac{1}{2}$)\cite{Kunth-Ostlin2000}, centrally concentrated \cite{hunter06} galaxies with bright clumps of star-formation \cite{Ostlinetal2021}. Cosmological surface brightness dimming\cite{Calvietal2014} and small size limit their detection at high redshifts, making their formation process difficult to observe. Observations of BCDs are needed at intermediate redshifts, where they are still young enough to show
their formative stages, particularly in the outer regions where cosmic gas accretion should drive evolution.
Here, we report the discovery of excess far-ultraviolet (FUV) emission in the outer regions of 11 BCDs in the GOODS-South field at redshifts between 0.1 and 0.24, corresponding to look back times of 1.3 - 2.8 Gyr in standard cosmology. These observations were made by the Ultra-Violet Imaging Telescope (UVIT)\cite{Tandonetal2017a} 
on AstroSat\cite{Singhetal2014}. For ten BCDs, the radial profiles of intrinsic FUV emission, corrected for the instrument point spread function, have larger scale-lengths than their optical counterparts observed with the Hubble Space Telescope. Such shallow FUV profiles suggest extended star-formation in cosmically accreting disks. Clumpy structure in the FUV also suggests the outer FUV disks are gravitationally unstable. Dynamical friction on the clumps drives them inward at an average rate exceeding $10^6~M_{\odot}$Gyr$^{-1}$.
}

\bigskip
\par

The formation and growth of BCDs are challenging to observe directly because these galaxies are too small to resolve at high redshift and too faint to reveal their outer regions in the Local Universe\cite{Hunteretal2011}. Far-ultraviolet observations at intermediate redshifts are necessary to show star-formation in their outer disks while they are still growing. For this purpose, we use a sample\cite{Lianetal2015} of 14 BCDs having deep observations by UVIT\cite{Tandonetal2017a} in the GOODS~South Field.
Two of the BCDs, GS3 ($z=0.1$) and GS6 ($z=0.21$), are shown in Figure 1, with the others in the Methods. Most have extended FUV (hereafter, XUV) emission beyond their optical disks, which are well resolved with the Hubble Space Telescope (HST). This is the first time such large FUV disks have been observed in distant dwarfs. There are also large FUV clumps of magnitude 25.3 (C3A), 25.46 (C3B) and 26.8 (C6A) in the XUV regions of GS3 and GS6 respectively. Nine clumps in the rest of the BCDs are listed in Table 4.
Most clumps were detected using automated software with the constraint of a minimum area of 5 pixels\cite{galametz2013}. Some clumps had to be separated from the main galaxy by eye (see Sect. 1.6.1 for clump detection significance). All magnitudes discussed in this paper are corrected for foreground extinction\cite{Schlafly-Finkbeiner2011} and K-correction\cite{boyleetal1998}.

\begin{figure}
    \centering
    \includegraphics[width=\textwidth]{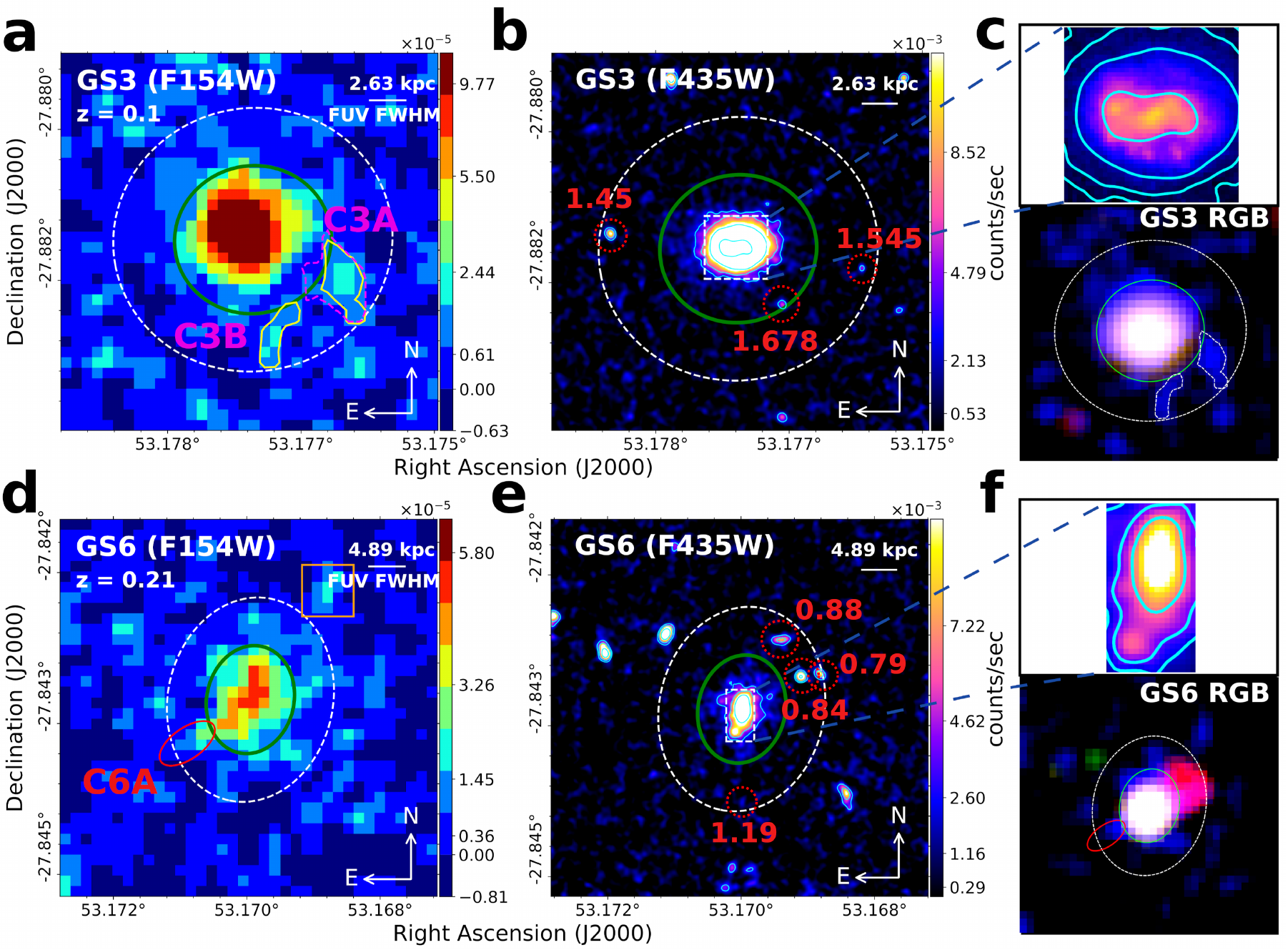}
    \caption{{\textbf{Detection of XUV disks in two BCDs}}. {\textbf{a, d}}: False color images of GS3 and GS6 in F154W band. Detected clumps C3A, C3B and C6A, present in the XUV regions of GS3 and GS6 respectively, have no optical counterparts. Clumps marked with magenta-dashed and yellow-solid contours were identified using automated tools, while the clump marked in the red ellipse (C6A) was identified visually. The source marked by the orange box (panel d) does not have an optical counterpart but is outside the FUV extent of GS6. $1$~pixel$\simeq $0.8 kpc and $\sim$1.4 kpc in GS3 and GS6, respectively. {\textbf{b,e}}: The contours in the HST/F435W optical images of GS3 and GS6 are drawn at 10, 20, 50, 250$\sigma$ ($\sigma_{GS3}$ = 3.1$\times$10$^{-4}$ and $\sigma_{GS6}$ = 2.9$\times$10$^{-4}$ counts/sec/pix). {\textbf{c,f}}: Top zoomed-in panels show the inner clumpy structures of GS3 and GS6 in the optical. The bottom panels are color composite images of the BCDs made using F154W band images combined with F435W and F160W  blurred by the FUV PSF. The red blob in the North-West of GS6 is composed of the background galaxies seen in the F435W image. For all images, the white dashed ellipse is the FUV extent down to S/N=3 in the measured 1D surface brightness profile and corrected for the width of the PSF by subtraction in quadrature (see Methods). The green ellipse is the optical extent after the HST image is convolved with FUV PSF (marked by solid line in each image) and similarly corrected by subtraction in quadrature. Red dashed circles in the HST images mark background galaxies with redshifts from the 3D-HST catalog\cite{Momchevaetal2016}.}
    \label{fig:fig1}
\end{figure}

Most disk galaxies including dwarfs follow intrinsic exponential profiles\cite{Herrmannetal2013}. The observed surface brightness profiles in different bands are modelled with exponential profiles convolved with the appropriate PSF (see Figure~2).
The fitted intrinsic profiles of GS3, GS6, and eight others (except GS12 and GS14) are shallower in the FUV than in the optical, implying the FUV disks are intrinsically more extended than the optical, not just apparently extended because of the broad PSF. 
FUV measurements were not possible due to non-detection in GS8; GS9 was too compact to model. For GS3 and GS6, the intrinsic FUV scale lengths are $\sim$1.4 and 2.4 times their respective optical scale-lengths (see Table~1). 

\begin{figure}
    \centering
    \includegraphics[width=\textwidth]{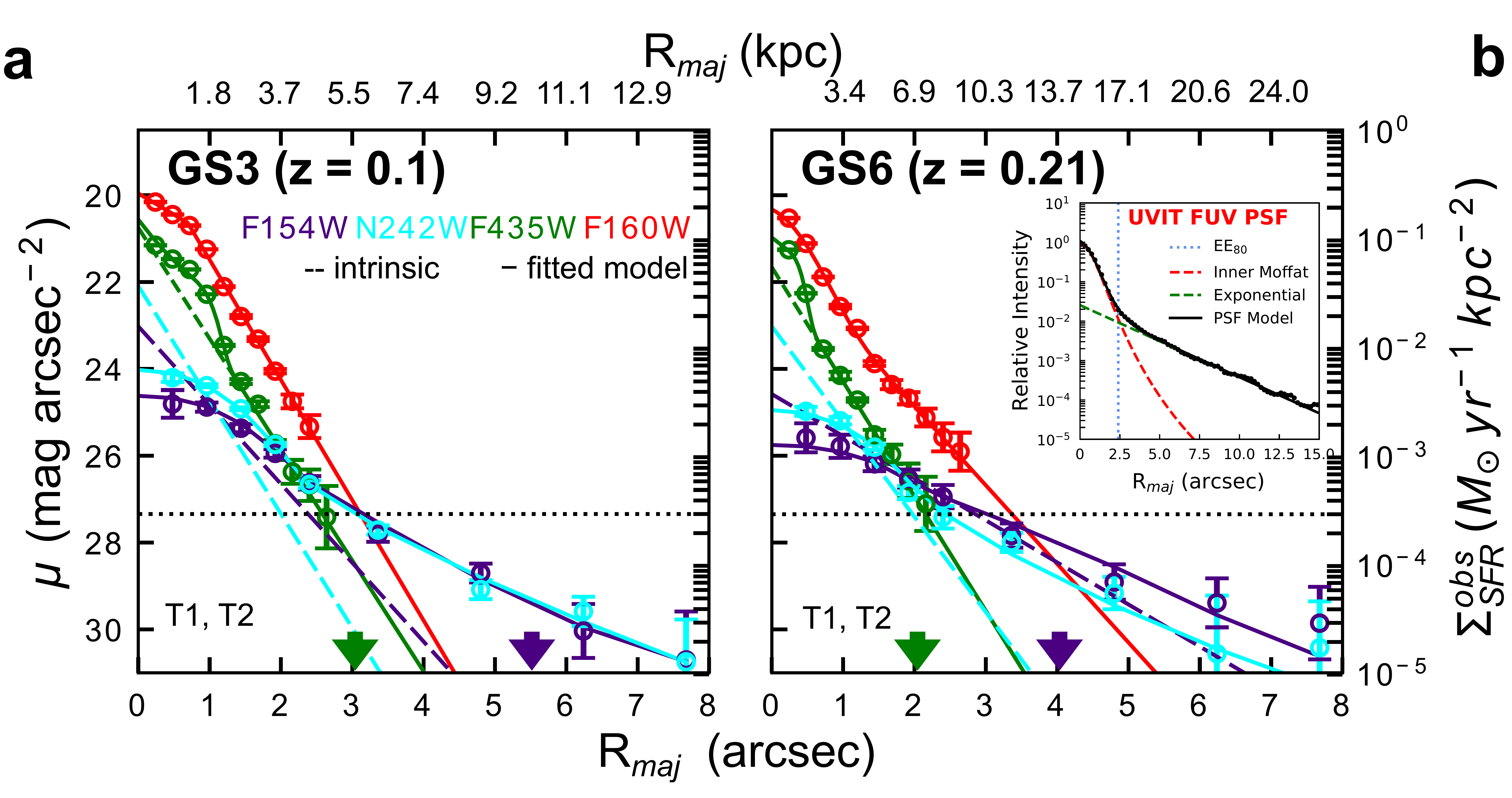}
    \caption{\textbf{1D Surface brightness profile fitting. a, b:} Solid curves show the intrinsic profiles convolved with the PSF as fits to the observed surface brightness. Circles are datapoints in F154W (indigo), N242W (cyan), F435W (green) and F160W (red) bands. The dashed curves in indigo (F154W), cyan (N242W) and green (F435W) represent the fitted intrinsic exponential profiles. The upside down arrows mark the F435W and F154W extents as shown in Figure~1. The horizontal dotted line marks the threshold surface brightness level for XUV disks\cite{Thilkeretal2007}. T1,T2 at the bottom left denotes the XUV disk type\cite{Thilkeretal2007}. The FUV PSF fit with an inner Moffat profile and an exponential wing is shown as an inset in panel \textbf{b}. All surface brightness data points have $1\sigma$ error bars.}
    \label{fig:fig2}
\end{figure}

Figure 2 shows that the observed FUV surface brightness, corrected for foreground extinction\cite{Schlafly-Finkbeiner2011} and cosmological dimming\cite{Calvietal2014}, goes at least as low as $\sim$28 - 29~ABmag~arcsec$^{-2}$. This is well below the threshold of $27.25$~ABmag~arcsec$^{-2}$ used as one of the criteria to define Type~1 XUV disks \cite{Thilkeretal2007} for local spirals. The equivalent threshold star-formation rate surface density (SFRD) with a Salpeter IMF and solar metallicity (Z$_{\odot}$=0.02)\cite{kennicutt12} is $SFRD_{th}=3 \times 10^{-4}$~M$_{\odot}$~yr$^{-1}$~kpc$^{-2}$ (see Methods). For the average intrinsic surface brightness in the XUV region of GS3 and GS6, the SFRDs are ${1.82} \pm {0.43}\times 10^{-5}$~M$_{\odot}$~yr$^{-1}$~kpc$^{-2}$, and ${2.25} \pm {0.48} \times 10^{-4}$~M$_{\odot}$~yr$^{-1}$~kpc$^{-2}$ respectively. These values would be $\sim 10$\% lower at 0.4Z$_{\odot}$ because the FUV emission of a 100 Myr old population is 1.1 times brighter\cite{hunter10}. Our SFRD estimates are comparable to previous measurements that reach $\sim$10$^{-5}$-10$^{-6}$ M$_{\odot}$yr$^{-1}$kpc$^{-2}$ (Figure~3a) in low-density, extreme environments, e.g., galaxy outskirts or stripped gas tails within galaxy clusters \cite{Hunteretal2011,jachym2014}.
If the FUV SFRD follows the Kennicutt relation \cite{kennicutt12}, it would suggest star-formation at an average gas surface density of less than $1 $~M$_{\odot}$~pc$^{-2}$ \cite{Zavadsky2014,Bicalhoetal2019}. For a steeper slope of the Kennicutt relation as seen in the XUV region of M63\cite{Zavadsky2014}, our SFRDs probe gas (HI+H$_{2}$) surface densities (independent of metallicity because the molecules are fully observed in M63) of $\sim$2.9 and $\sim$4.9 M$_{\odot}$pc$^{-2}$ for GS3 and GS6 respectively. The nature of star-formation at such low-density is largely unexplored.

\begin{figure}
    \centering
    \includegraphics[width=\textwidth]{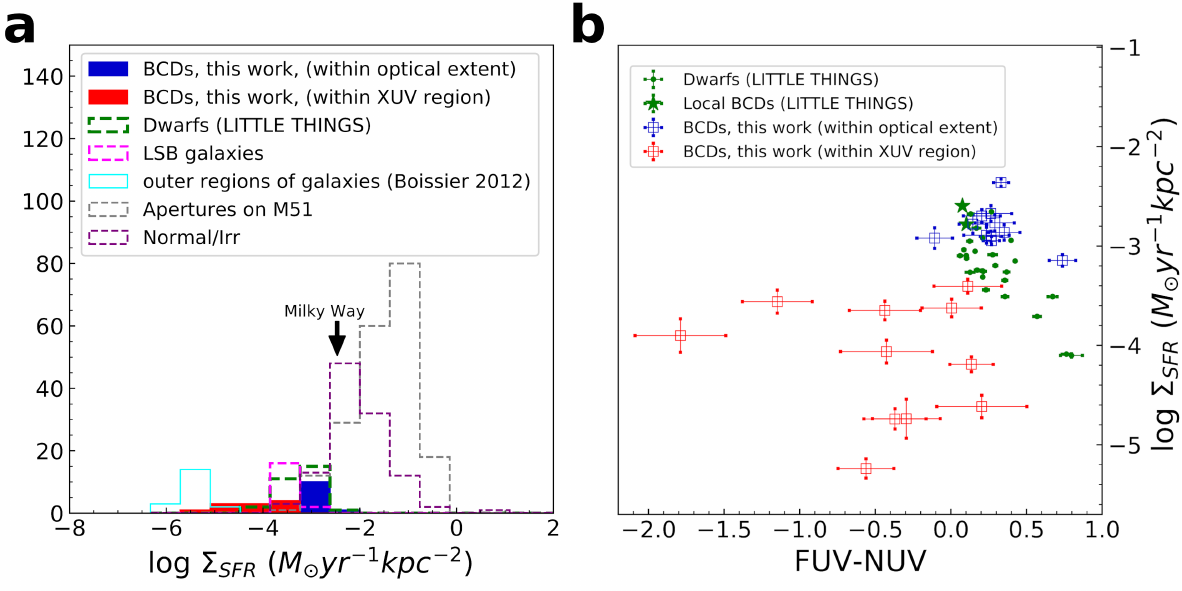}
    \caption{\textbf{Comparison of star-formation properties of our BCDs to local galaxies:} \textbf{a:} SFRD values of our sample BCDs (shown in filled blue and red), other local dwarfs (dashed-green -- measured with GALEX within their Holmberg radii\cite{hunter12}), outer regions of galaxies\cite{boissieretal2012} (solid-cyan), low surface brightness (LSB) galaxies, apertures on M51 and other Normal/Irr galaxies\cite{kennicutt12}. \textbf{b}: SFRD versus observed FUV-NUV color of the current BCD sample (blue and red squares),  local BCDs (green-stars) and other local dwarfs\cite{hunter12} (green points). The measurements for local BCDs and other dwarfs are within their Holmberg radii\cite{hunter12}. All our measurements are shown with 1$\sigma$ errorbars.}
    \label{fig:fig3}
\end{figure}

Figure~3b shows the SFRD versus $FUV-NUV$ colors, directly measured from the extinction corrected UVIT images, for 11 of our BCDs (excluding GS5). 
Since massive OB stars typically dominate far-UV emission, FUV and NUV can detect star-formation in the past $100$ to $200$~Myr.
Negative values of the color indicate very young portions of the disk. Seven of our galaxies have $FUV-NUV <0$ while four have $0<FUV-NUV < 0.5$ in their XUV region. GS3 and GS6 have $FUV-NUV \sim 0.3 $ and $0.2$ respectively, within the observed optical extent, similar to local LITTLE THINGS dwarfs\cite{hunter12}. We calculate the stellar population ages of these XUV disks by comparing their observed $FUV-NUV$ colors with model colors obtained using the STARBURST99 code\cite{leitherer99} (Methods). In the case of an instantaneous burst at Z$_{\odot}$, the ages are 7.1 Myr and 12.3 Myr for GS3 and GS6, respectively, whereas for a continuous star-formation history, they are 106 Myr and 306 Myr old (the stellar populations are older by a factor of $\sim2$ at 0.4~Z$_{\odot}$).
\par

The relative excess FUV emission beyond the convolved optical extent, denoted by $\Delta L_x$, for 12 BCDs is shown in Figure~4. Ten of them have $R_{h,{FUV}}/R_{h,{opt}} > 1$ and $\Delta {L_{x}}>0$, including GS5, the marginal case. GS12 and GS14 have $R_{h,{FUV}}/R_{h,{opt}} < 1$ but $\Delta {L_{x}}>0$. These parameters can be useful to infer the presence of XUV emission when the spatial resolution is poor, especially in distant galaxies. In our downsized sample of 12 BCDs, 7 with clumps qualify for Thilker's Type-1 XUV criteria, 10 follow the Type-2 criteria, and 5 follow both types \cite{Thilkeretal2007}(see Methods and Table~5).

\begin{figure}
    \centering
    \includegraphics[width=0.5\textwidth]{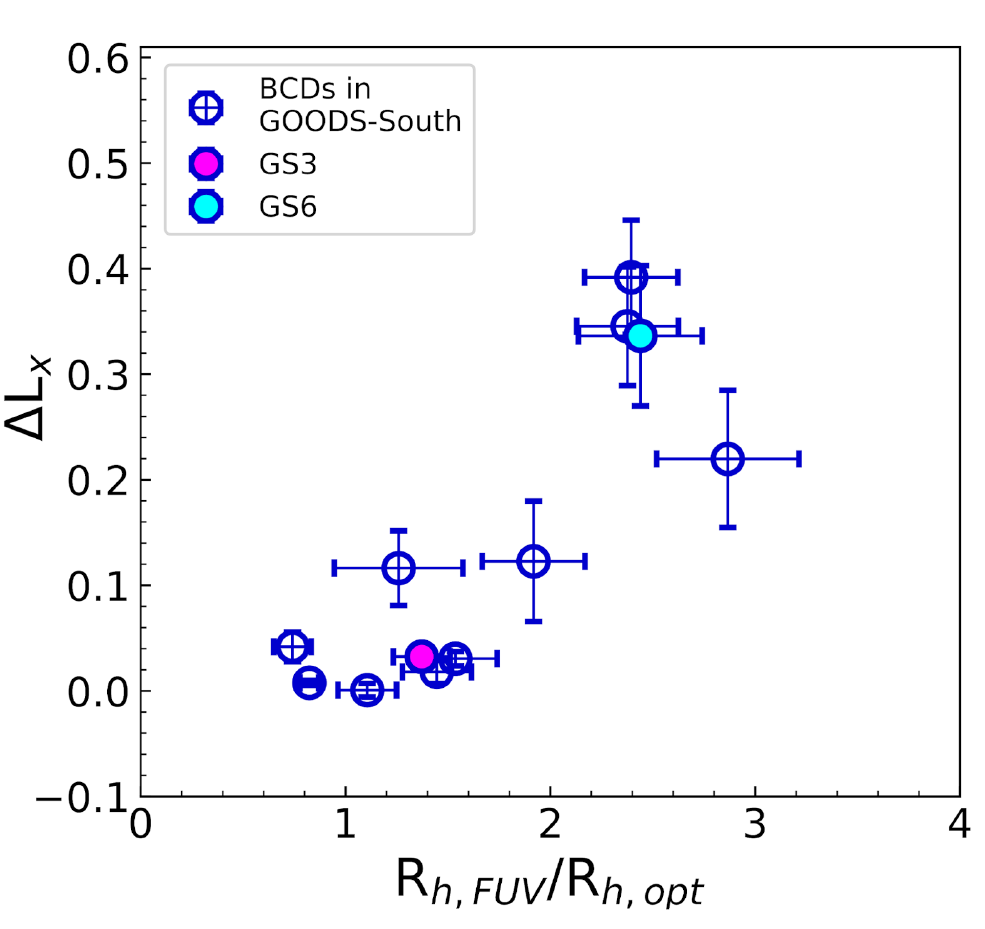}
    \caption{\textbf{Relation between excess UV light and disk scale lengths.} The fraction of the total FUV light that is in the XUV disk (defined by the annulus between the optical and FUV extents), $\Delta L_x$, is plotted versus the ratio of the intrinsic, PSF-corrected, FUV to optical scale length along with 1$\sigma$ errorbars.}
    \label{fig:fig4}
\end{figure}

\par
XUV emission implies a significant fraction of young stars at large radii, possibly accompanied by an ongoing assembly of the outer disk over Gyr timescales. We assess the outer disk evolution from its clumpy structure, focusing on seven BCDs resembling star-forming disks of high-redshift massive galaxies\cite{elmegreen2009}. As resolved by HST, the optical disks are often clumpy too (insets in Figure~1 c,f). The clumpy structure suggests the disks are gravitationally unstable, and the clump size is comparable to the turbulent Jeans length. For GS3, the large FUV clumps have SFR of a few $\times10^{-2}\;M_\odot$ yr$^{-1}$. Considering a $\sim100$ Myr timescale of FUV emission, their masses would be a few $\times10^6\;M_\odot$ (Table 4). This is 1\% of the galaxy's total stellar mass, $M_{*}=3.70\times10^8\;M_\odot$ (Table~5). The stellar masses of the optical clumps are similar, $\log M_{*}/M_{\odot} = 6.95$ and 6.33, based on their optical colors and brightnesses\cite{bell01}. Such a clumpy structure suggests that strong torques may drive in-plane accretion.

\par
The timescale for in-plane accretion of a clump may be determined as follows\citep{Elmegreenetal2012}. The instantaneous dynamical friction force \cite{binney08} (${\vec{F}_{df}}$) acting on an orbiting clump of mass $M_{\rm c}$ from the surrounding dark matter halo is given by 

\begin{equation} 
    {\vec{F}_{df}}= -{{ 4 \pi \ln\Lambda G^2M_{\rm c}^2
    \rho}\over V^2} \left( {\rm erf}\left[X\right] - {{ 2X}\over{\pi^{1/2}}}e^{-X^2} \right) {\hat{V}}=-{{ 4 \pi \ln\Lambda G^2M_{\rm c}^2 \rho\xi}\over V^2}{\hat{V}}; 
    \label{eq:dvdt}
\end{equation}
 
\noindent $V$ is the clump orbital speed, ${\hat{V}}$ is the unit vector along the velocity direction, $X=V/(2^{1/2}\sigma)$ for halo 3D velocity dispersion $\sigma$ with a Maxwellian distribution function, $\rho$ is the halo density and $\xi$ is the quantity in parentheses. The $\xi$ term increases from 0 to 1 as $X$ increases. We assume a pseudo-isothermal halo \cite{binney08} with a core radius $R_c$ and associated asymptotic rotation speed $V_{0} = \sqrt{2} \sigma$. Assuming a factor of $\alpha = 2.7$ to convert stellar mass to dynamical mass as in the local BCD, NGC 4861\cite{Elmegreenetal2012}, we obtain the dark matter halo mass, $M_{dm}$=$6.42\times10^{8}$~M$_{\odot}$ for GS3. Then the circular velocity at the radius of the clump C3A (Fig.1) is $V_{c0} = 25$~km s$^{-1}$, and the Coulomb factor is $\ln \Lambda \simeq 5$. The in-spiral timescale, $T_{insprial}$, is then derived by integrating the equation for angular momentum change, $d\vec{L_{z}}/dt = \vec{r} \times \vec{F}_{df}$, from the current clump position to the optical radius, $R_{opt}$ for all $12$ clumps detected in our sample of BCDs (see  Table 4). For the largest clump C3A in GS3, $T_{inspiral}\sim 5.6$~Gyr is $4.3$ times the look-back time at z=0.1. For the clump in GS14 (z=0.1), C14A, $T_{inspiral}\sim 2.4$~Gyr.

\par
The clump mass divided by $T_{inspiral}$ gives the instantaneous clump accretion rate, and their sum for each galaxy gives the summed clump accretion rate, $\dot{M}_{clump}$. The average value of $\dot{M}_{clump}$ for all the galaxies is $\sim1.1\times10^{6}$~M$_{\odot}$~Gyr$^{-1}$. For GS3, $\dot{M}_{clump}\sim2.3\times 10^{6}$~M$_{\odot}$~Gyr$^{-1}$ and the average clump accretion time is $\tau_{clump}=5.5$~Gyr. This time is less than Hubble time, implying that clump torques can lead to significant evolution of the outer structures. 
\par
The time, $\tau_{XUV}$, over which a mass equal to all the young stellar mass in the XUV disk would be accreted by clump torques is larger than $T_{inspiral}$ because the XUV mass is larger than the summed clump mass. $\tau_{XUV}$ is several Hubble times for most BCDs (except for GS3: 6.6 Gyr and GS14: 3.7 Gyr; Table~5). 
The accretion time for the additional old stars in the outer disk would be longer still.
Interestingly, $T_{inspiral}$ is comparable to the galactic star formation time for our BCDs, $\tau_{SF}=M_{*,total}/SFR$ for total stellar mass from Table 5, $M_{*,total}$, and FUV SFR from Table 1.
The relatively short $\tau_{SF} \sim$~few Gyr - implies a rapid buildup of the main disk at the same time as the outer disk clumps move inward.

\par
The accretion times should be considered upper limits because the young stellar clumps are more massive if they contain gas. Since typical star-formation efficiencies in the youngest regions are only a few percent\cite{kennicutt12}, the mass multiplier from gas could be a factor of 50 for such regions, decreasing the timescale in inverse proportion. The timescale would also be lower than just from the halo torques because disk stars and gas may also exert torques on a clump \cite{Ceverinoetal2010}. In addition, differential shear in the outer disks could bring the clumps closer together so they coalesce and form bigger clumps with faster accretion. The clump collision time from shear ($\tau_{cc}$ in Table~4) is approximately the orbit time. If these effects are considered, the clump accretion time could be shorter by a factor of 5 or more. Then for BCDs with $\Delta L_{x} \ge 10\%$, the effective accretion time for the young stellar mass in the XUV disk, $\tau_{XUV} \sim$~a few Gyr. In GS3 and GS14 with this revision, $\tau_{XUV}\sim 1$ Gyr. Such large young stellar accretion rates are likely to be accompanied by simultaneous gas accretion. 

We conclude that the BCDs studied here, at redshifts of 0.1 to 0.24, typically have more extended FUV disks than their optical counterparts, suggesting significant outer, low-density star formation. These extended FUV disks contain $10^6\;M_\odot$ clumps that produce enough torques to drive them, or an equivalent mass, inward to the optical disk, contributing to a more centrally concentrated structure. The torques are not large enough to bring in the whole outer disks, which should fade into the extended old disks\cite{Zhangetal2012} and halos\cite{Schulte-Ladbecketal1999} of today's BCDs\cite{Ostlinetal2021}.

\bigskip
\bigskip

\medskip{}
\noindent {\bf Acknowledgement}\\
This publication uses the data from the AstroSat mission of the Indian Space Research Organisation (ISRO), archived at the Indian Space Science Data Centre (ISSDC). The far and near-UV observations were carried out by UVIT which is built in collaboration between IIA, IUCAA, TIFR, ISRO and CSA. A.B. acknowledge the support of the DST-INSPIRE Fellowship program by the Government of India. A.B. and R.G. would like to thank the IUCAA associateship programme for their support and hospitality. K.S. and F.C. acknowledge the support of CEFIPRA-IFCPAR grant through the project no. 5804-1 during its initial phase.

\medskip{}

\medskip{}
\noindent {\bf Corresponding author}\\
\noindent Correspondence to Kanak Saha  (Email: kanak@iucaa.in)\\
\medskip{}

\noindent {\bf Ethics Declarations}\\
\noindent {\large Competing interests}\\

\noindent The Authors declare no competing interests.

\newpage
\section{Methods}
\par
A flat $\Lambda$~Cold Dark Matter cosmology with Hubble constant, $H_{0}$ = 70 km$s^{-1}$ Mpc$^{-1}$ , $\Omega_{m}$ = 0.3, and $\Omega_{\Lambda}$ = 0.7 was adopted throughout the article. All magnitudes quoted hereafter in the paper are in the AB system\cite{oke1974}.

\subsection{AstroSat observation and other archival data}
\label{data_descrip}
\par
Our sample of BCDs is selected based on the following criteria - (i) central surface brightness $< 22$~mag~arcsec$^{-2}$ in HST F435 filter; (ii) rest-frame $B-R < 1.0$; (iii) stellar mass $<10^{9}$~M$_{\odot}$ and redshift $z < 0.25$ (obtained from Skelton et al.\cite{Skeltonetal2014}) considering surface brightness dimming affecting the observation of low-surface brightness outskirts of these galaxies. These criteria select only 14 BCDs \cite{Lianetal2015} in the GOODS-South field for which  high-quality multi-wavelength data are available.  

In the current study, we use archival observations from HST \citep{Illingworthetal2016} and HAWK-I \citep{Fontanaetal2014} along with new Ultraviolet (UV) observations from UVIT\cite{Tandonetal2017a}. The orbit-wise UVIT dataset\cite{Sahaetal2020} of GOODS-South field (P.I. Kanak Saha, ID: GT05-240) was processed using the official L2 pipeline. We removed frames affected by the cosmic-ray shower. These frames were excluded (resulting in $\sim 15\%$ data-loss) in the final science-ready images and the subsequent photometry calculation. In addition, there was data loss due to the mismatch in time-stamp on the VIS (visual) filter and NUV or FUV filters. The final science-ready images had a total exposure time of about 17.7 and about 17.3 h in F154W and N242W, respectively \cite{Sahaetal2020}. The PSF FWHMs are $\sim$1.4 arcsec (F154W) and 1.3 arcsec (N242W) - more than a factor of 3 better resolution than the GALEX UV deep field \cite{Xuetal2005}.

\par
We estimate the sky background following a standard procedure. We run \textit{Source Extractor}\cite{Bertin_Arnouts1996} (\textit{SExtractor}) on the parent UVIT image with a low detection threshold and dilate the obtained segmentation map to mask all detected sources. We then placed a large set of boxes of size 9 pixels randomly over the masked image and created a flux (within the box) histogram. The procedure also required visual inspection to avoiding the masked regions. The resulting histogram was fitted with a Gaussian function whose mean is the estimated background. For FUV and NUV, they are $28.43\pm0.01$~mag~arcsec$^{-2}$ and $27.6\pm 0.003$~mag~arcsec$^{-2}$ respectively. The 3$\sigma$ (5$\sigma$) limiting point source magnitude within a aperture of radius 1" is $\sim$27.94 (27.44) mag and $\sim$28.17 (27.62) mag in the FUV and NUV respectively. We followed the same procedure but on a 101$\times$101~pix cutout to estimate the local background around each BCD. We do this because the local sky can differ at two different locations. This can affect the photometry of fainter structures in the outer regions. For the BCDs in our sample, the galaxy IDs have been kept same as in literature\cite{Lianetal2015} for comparison. 

\par
All measured magnitudes have been corrected for Galactic extinction\cite{Schlafly-Finkbeiner2011} using a E(B$-$V) = 0.008 mag towards the GOODS-South field\cite{Schlegeletal1998}. We use $\frac{A_{\lambda}}{E(B-V)}$ = 8.06 and 7.95 for FUV and NUV passbands respectively\cite{Bianchi2011}.

\begin{figure}[!ht]
    \centering
    \includegraphics[width=\textwidth]{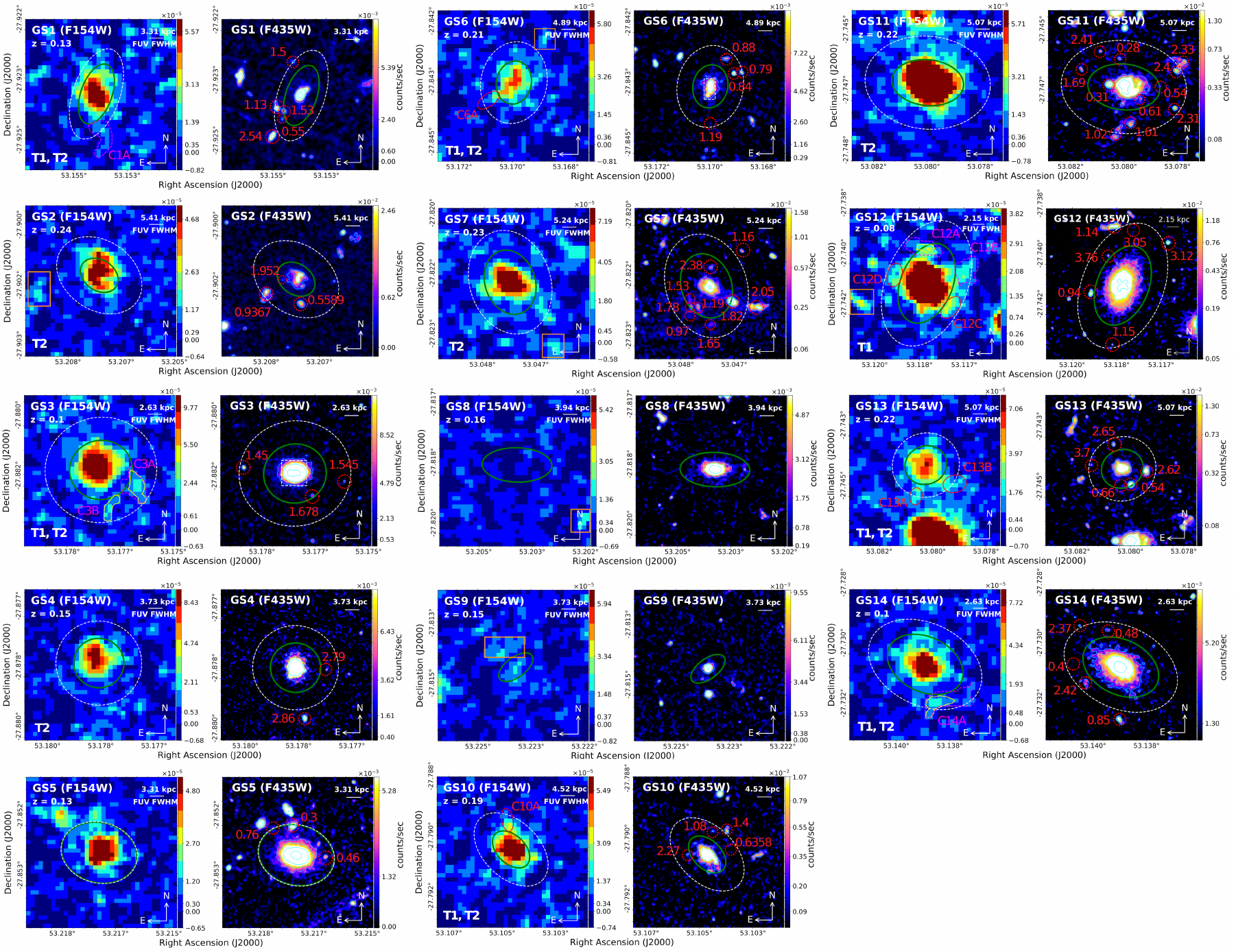}
    \caption{\textbf{Imaging of BCD sample. }Same as  Fig.1 (a,b,d,e) but for the rest of the BCDs in our sample. Individual panel sizes are 15''x15'', except GS12--18''x18''. We include the XUV disk type, if qualified, for each BCD in the lower left corner of each panel. We find a total of 6 sources (orange boxes) having S/N$\geq$3 and without any HST counterpart outside the XUV region of the BCDs.}
    \label{fig:fig5}
\end{figure}

\subsection{Multiband surface photometry}
We fit elliptical isophotes to the BCDs in background-subtracted F435W band images using the IRAF\cite{Tody1986} ELLIPSE\cite{Jedrzejewski1987} task. We mask all the sources around each BCD prior to performing surface photometry. While fitting isophotes in F435W imaging, the center was kept fixed. We then obtain the surface brightness profiles (SBPs) as follows. The intensity ($I$ in units of counts/s/pix) at a semi-major axis (SMA) length $r$ is obtained by measuring the flux within an annulus of SMA  $r$ and $r+\Delta$r and dividing it by the total number of pixels in that annulus. Here, $\Delta r =4$ pixels $=0.24$'' in HST imaging. After correcting the intensity ($I$) for redshift dimming by a factor of (1+z)$^4$, the surface brightness ($\mu$) at a given SMA is given by

\begin{eqnarray}
    \mu = -2.5\: \log \frac{I}{pixel\:scale^2} + magZP
\end{eqnarray}

The isophotes fitted to the F435W imaging are placed on the F154W, N242W and F160W images to obtain the SBPs in respective passbands. The $\Delta$r, however, ranges from 0.48'' (inner regions) to 1.44'' (outer parts) so that a sub-pixel sampling in UV is avoided in the inner parts (1 pixel in UVIT = 0.417'') and at the same time have a large enough area in the outer parts (comparable to FUV PSF $\sim$1.4'') to collect sufficient signal that have been affected by the FUV PSF.

\par
The photometric zeropoints\citep{Tandonetal2017a,Illingworthetal2016} are given by, magZP = 17.78 (F154W), 19.81 (N242W), 25.68 (F435W), and 25.94 (F160W). The error in the surface brightness values ($\Delta \mu$), measured for the total flux within each annulus area, is estimated by taking into account the background sky rms within each annulus as follows:

\begin{eqnarray}
    flux\:error = \sqrt{n_{pix}\sigma_{sky}^{2} + \frac{flux}{exptime}} \: \: \: and \: \: \: \Delta\mu = \frac{flux\:error}{flux} \frac{2.5}{\ln{10}}
\end{eqnarray}

\noindent where $n_{pix}$ is the number of pixels within each annulus, flux and exptime refer to the total counts/s within the annulus and total exposure time respectively. For the HST images, the associated weight images are used to compute the sky rms, $\sigma_{sky}$. The median values of locally obtained $\sigma_{sky}$ are 6.41$\times10^{-6}$ and 3.54$\times10^{-5}$ counts/s/pix in the F154W and N242W respectively. These median values correspond to limiting surface brightness levels of 29.6 (FUV) and 29.7 (NUV) mag arcsec$^{-2}$ at 3$\sigma$ above the background in a circular area of radius equal to PSF FWHM. 

\noindent {\bf Outer extent of a galaxy:}\\
Throughout our analysis, along with the 1D SBP, we also obtain 1D S/N profile for each galaxy and mark their extents in any band as the extent beyond which the S/N goes below 3 (this position is determined by a combination of the galaxy profile and the sky noise and would be a larger extent for lower sky noise). The FUV and optical extents (See Figure~1) are corrected for the effect of the PSF as follows: 

\begin{equation}
    r_{corr} = \sqrt{r_{obs}^{2} - hwhm^{2}}
\end{equation}
where r$_{corr}$ is the corrected extent, r$_{obs}$ is the observed extent at which we find S/N=3, and hwhm is the half width half maximum in a filter. Unless otherwise noted, all the extents quoted in the text have been corrected for the PSF as above.

\subsection{Modelling of surface brightness profiles}
\label{sec:profile-fitting}
 We use \emph{PROFILER}\cite{Ciambur2016} to model the HST SB-profiles with a combination of Sersic plus exponential or an exponential alone\cite{Lianetal2015}. The disk SB-profiles can be broadly classified into Type-I, Type-II and Type-III\cite{Erwinetal2008}. So, in addition to Sersic plus exponential, we also consider truncated exponential profile for a more realistic modelling of the HST images of the BCDs (see Figure~2).

\begin{figure}[!ht]
    \centering
    \includegraphics[width=0.8\textwidth]{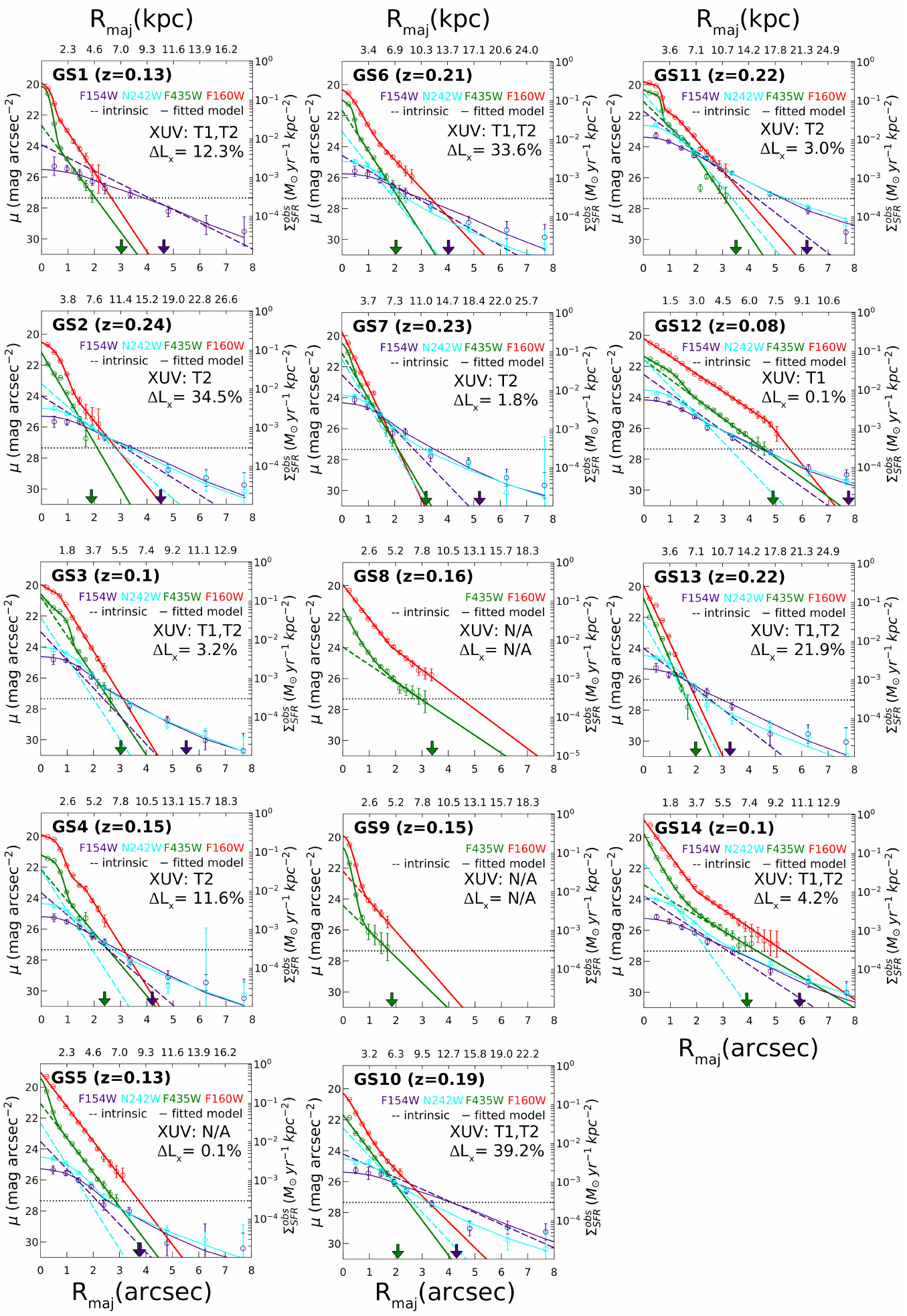}
    \caption{\textbf{1D Surface brightness profile fitting of BCD sample.} Same as in Fig.~2 but for the rest of the BCDs in our sample. In each panel, we include the XUV types and $\Delta$L$_{x}$ values for each BCD. All data points have $1\sigma$ error bars.}
    \label{fig:fig6}
\end{figure}

\begin{table}[!ht]
\begin{threeparttable}
\normalsize{
\centering
\caption{\textbf{Photometric and structural parameters of the BCDs.} Column 1: The galaxy IDs. Column 2: The redshifts. Column 3: The K$_{s}$ and FUV band absolute magnitudes. Column 4 and 5: The central surface brightness values in all four bands - FUV (F154W), NUV (N242W), optical (F435W) and NIR (F160W), corrected for redshift dimming and foreground extinction. Column 6 and 7: The scale-lengths from FUV to NIR; for comparison purposes, if a truncated exponential is used, the outer scale-length is given here. Column 8: The measured total SFRs from the FUV light assuming solar metallicity (these rates are lower by 10\% at 0.4Z$_{\odot}$). Column 9: The measured FUV - NUV colors. All quoted uncertainties are 1$\sigma$.  }
\label{table:SBPs}
\begin{tabular}{c c c c c c c c c c c }
\specialrule{0.75pt}{0pt}{0pt}
BCD		&	 z		&		M$_{FUV}$		&	$\mu_{0,\:FUV}$ 	&		$\mu_{0,\:opt}$		&		\textbf{r$_{h,FUV}$}		&		\textbf{r$_{h,opt}$}		&	   SFR 	&	   FUV-NUV\\
&				&	M$_{K}$ 	&	$\mu_{0,\:NUV}$	&	$\mu_{0,\:nir}$ 	&		r$_{h,NUV}$		&	 r$_{h,nir}$ 	&	   	&	\\
\specialrule{0.75pt}{0pt}{0pt}														
&				&		mag 		&		mag arcsec$^{-2}$ 		&		mag arcsec$^{-2}$ 		&		kpc 		&		kpc 		&	  M$_{\odot}$yr$^{-1}$    	&	mag   \\
\specialrule{0.75pt}{0pt}{0pt}																																			
GS1 		&		 0.13				&			-15.13				&		23.9$\pm$0.3			&			22.7$\pm$0.2			&	\textbf{2.1$\pm$0.3}  	&	\textbf{1.1$\pm$0.0} 	&	 0.08$\pm$0.00  	&	-0.32$\pm$0.11   \\
&			&	   	-16.97		&		--		&			20.5$\pm$0.1			&		--		&			1.0$\pm$0.0		&		  	&		\\
\specialrule{0.75pt}{0pt}{0pt}

GS2 		&		 0.24				&			-16.40				&			24.0$\pm$0.2			&			21.2$\pm$0.1			&			\textbf{3.4$\pm$0.3}  	&			\textbf{1.4$\pm$0.1}		&	 0.30$\pm$0.02 	&	 0.17$\pm$0.09  \\
&						&			-18.44				&			23.2$\pm$0.1			&			21.4$\pm$0.6			&			2.4$\pm$0.1		&			2.0$\pm$0.2		&			&		\\
\specialrule{0.75pt}{0pt}{0pt}																														
GS3			&		 0.1			&			-15.47			&			23.0$\pm$0.1			&			20.7$\pm$0.3			&			\textbf{1.1$\pm$0.1}  	&			\textbf{0.8$\pm$0.1}  	&	   0.11$\pm$0.00   	&	   0.18$\pm$0.05\\
&						&			-18.05			&			22.1$\pm$0.1			&			20.0$\pm$0.1			&			0.7$\pm$0.0   	&			0.7$\pm$0.0  	&	   	&	\\
\specialrule{0.75pt}{0pt}{0pt}																														
GS4			&		 0.15			&			-15.57			&			23.7$\pm$0.2			&			22.2$\pm$0.8			&			\textbf{1.8$\pm$0.2}  	&			\textbf{1.4$\pm$0.3}  	&	   0.13$\pm$0.01   	&	   0.25$\pm$0.08\\
&						&			-18.81			&			22.2$\pm$0.1			&			20.9$\pm$0.9			&			1.0$\pm$0.1   	&			1.1$\pm$0.0  	&	   	&	\\
\specialrule{0.75pt}{0pt}{0pt}																							
GS5			&		 0.13			&			-15.00			&			23.5$\pm$0.3			&			21.1$\pm$0.1			&			\textbf{1.3$\pm$0.2}  	&			\textbf{1.1$\pm$0.0}  	&	   0.07$\pm$0.00   	&	   0.34$\pm$0.09\\
&						&			-19.04			&			22.3$\pm$0.1			&			19.0$\pm$0.1			&			0.8$\pm$0.1   	&			1.1$\pm$0.0  	&	   	&	\\
\specialrule{0.75pt}{0pt}{0pt} 																
GS6			&		 0.21			&		-15.87				&			24.6$\pm$0.2			&			21.7$\pm$0.3			&			\textbf{3.4$\pm$0.3}  	&			\textbf{1.4$\pm$0.1}  	&	   0.18$\pm$0.01   	&	   0.04$\pm$0.11\\
&						&			-18.48			&			23.0$\pm$0.1			&			21.3$\pm$0.3			&			1.5$\pm$0.1   	&			2.1$\pm$0.1  	&	   	&	\\
\specialrule{0.75pt}{0pt}{0pt}																	
GS7 		&		 0.23				&		-16.91				&			22.5$\pm$0.2		&			21.1$\pm$0.1			&			\textbf{2.0$\pm$0.2}  	&			\textbf{1.4$\pm$0.0}		&	 0.47$\pm$0.02 	&	0.07$\pm$0.09   \\
&						&		-18.18			&			21.5$\pm$0.1		&			19.7$\pm$0.1			&		1.2$\pm$0.1		&			1.1$\pm$0.0		&		    	&		\\
\specialrule{0.75pt}{0pt}{0pt}																	

GS8 		&		 0.16				&						&				& 23.96$\pm$0.31						&			  	&		2.23$\pm$0.23		&	 	&   \\
&						&		-18.30		&			&	19.97$\pm$0.07					&				&		2.35$\pm$0.14		&		    	&		\\
\specialrule{0.75pt}{0pt}{0pt}																

GS9 		&		 0.15				&						&				&24.43$\pm$0.55						&			  	&		1.70$\pm$0.46		&	 	&   \\
&						&		-17.07		&				&		22.18$\pm$0.08			&				&		1.46$\pm$0.04		&		    	&		\\
\specialrule{0.75pt}{0pt}{0pt}																

GS10		&		 0.19				&			-15.92			&			24.2$\pm$0.2			&			21.7$\pm$0.1			&			\textbf{3.6$\pm$0.3}  	&			\textbf{1.5$\pm$0.1}  	&	   0.18$\pm$0.01   	&	   0.18$\pm$0.09\\
&						&			-18.03			&			22.6$\pm$0.1			&			21.9$\pm$0.8			&			1.5$\pm$0.1   	&			2.1$\pm$0.3  	&	      	&	\\
\specialrule{0.75pt}{0pt}{0pt}																														
GS11 		&		 0.22				&			-17.96			&			21.7$\pm$0.1			&			21.1$\pm0$.8			&			\textbf{2.4$\pm$0.1}  	&			\textbf{1.5$\pm$0.2}  	&	   1.24$\pm$0.03   	&	   0.31$\pm$0.04\\
&						&			-19.66		&			20.5$\pm$0.1			&			21.7$\pm0$.8			&			1.5$\pm$0.0   	&			2.0$\pm$0.0  	&	   	&	\\
\specialrule{0.75pt}{0pt}{0pt}																														
GS12 		&		 0.08				&			-15.53			&		 	22.5$\pm$0.1			&			21.6$\pm$0.1			&			\textbf{1.1$\pm$0.1}  	&			\textbf{1.3$\pm$0.0}  	&	   0.11$\pm$0.00   	&   0.19$\pm$0.04\\
&						&			-18.43			&			21.6$\pm$0.1			&			20.2$\pm$0.0			&			0.7$\pm$0.0   	&			0.8$\pm$0.0  	&	   	&	\\
\specialrule{0.75pt}{0pt}{0pt}																														
GS13 		&		 0.22				&			-16.19			&			24.0$\pm$0.2			&			20.8$\pm$0.2			&			\textbf{2.8$\pm$0.3}  	&			\textbf{1.0$\pm$0.1}  	&	   0.24$\pm$0.01   	&	   0.00$\pm$0.09\\
&						&			-18.13			&			22.3$\pm$0.2			&			19.9$\pm$0.2			&			1.2$\pm$0.1   	&			1.1$\pm$0.0  	&	   	&	\\
\specialrule{0.75pt}{0pt}{0pt}																														
GS14	&	0.1	&	-14.99	&	 \hspace{2mm}23.8$\pm$0.2 	&			23.1$\pm$0.3 	&	 \hspace{2mm} \textbf{1.5$\pm$0.1} 	&	    \textbf{2.0$\pm$0.2} 	&	   0.07$\pm$0.00   	&	   0.68$\pm$0.07\\[-0.2ex]
&	  	&	-18.62	&	 \hspace{2mm}21.7$\pm$0.1 	&			18.9$\pm$0.1 	&	 \hspace{2mm} 0.7$\pm$0.0   	&	    1.7$\pm$0.1 	&	   	&	\\[-0.2ex]
\specialrule{0.75pt}{0pt}{0pt}
\end{tabular}
}

\end{threeparttable}

\end{table}

 \par The UV disk was modelled with a different approach because errorbars are not taken into consideration in \emph{PROFILER}\cite{Ciambur2016}. For simplicity, we consider a pure exponential disk in both FUV and NUV having the same ellipticity as that of the outermost fitted ellipse (having S/N $\geq$ 3) in the observed HST/F435W. We do not rule out the existence of  truncated exponential disks, but given the resolution and angular size of the galaxies, we cannot justify assuming this. Thus the intrinsic UV disks are assumed to be single exponential. By convolving the exponential disk with the UVIT PSFs, we create model disk galaxies in UV. We obtain one-dimensional (1D) SBPs of the models at the same SMA lengths as considered for UV surface photometry. The 1D model profiles are then fitted to the observed UV profiles via least-squares optimization using the MPFIT routine\cite{MPFIT}, which is an implementation of the MINPACK\cite{More1981} Levenberg-Marquardt nonlinear least-square optimization algorithm. This provides us with the deconvolved scale-lengths and central surface brightnesses of the BCDs in the UV. The method can effectively recover an intrinsic exponential disk of a galaxy. However, multiple intrinsic components cannot be effectively recovered and the outcome could be biased. Photometry in faint out-skirts of galaxies is also sensitive to the sky background. To test the effect, we model the profiles again by subtracting sky values $\pm$1$\sigma$ about the mean sky. We also include this sky uncertainty into the error budget of the radial profiles. It is found that for nine BCDs, a change in $\pm$1$\sigma$ sky results in a change of FUV scale length up to 20\%. In cases of GS7, GS13 and GS14, the difference was $70$\%.

\subsection{Parameters of the XUV disks}

We check for the XUV emission in the BCDs using the classification scheme defined for the local galaxies\cite{Thilkeretal2007}. We convert the surface brightness values into the equivalent SFRDs following the SFR calibration\cite{kennicutt12} at solar metallicity as:

\begin{equation}
\label{sfrd}
\scalebox{1.4}{
    $SFRD = \frac{1}{A^{2}} \times 10^{\frac{\mu + 48.59}{-2.5}} \times 4 \pi D_{L}^{2} \times 1.4 \times 10^{-28}$},
\end{equation}

\noindent where SFRD is in $M_{\odot}~yr^{-1}~kpc^{-2}$, $\mu$ is the surface brightness in mag arcsec$^{-2}$, A is the angular scale in kpc/arcsec and D$_{L}$ is the luminosity distance in cm. The presence of structured UV emission beyond a threshold value of $SFRD_{th} =  3 \times 10^{-4}$~ $M_{\odot}~yr^{-1}~kpc^{-2}$ $\equiv$ 27.25~mag arcsec$^{-2}$ defines a Type~1 XUV disk\cite{Thilkeretal2007}. Sizes of such structures are considered to be $\sim 1$~kpc in Thilker's definition\cite{Thilkeretal2007}. However, at $z \sim 0.1 - 0.24$, the size of a minimum 4-connected pixels in the FUV image will be in the range of $\sim 1.5 - 3$~kpc. Although it is strictly not possible to follow the Type-1 definition at these redshifts, the presence of star-forming clumps beyond the detectable optical radii (Fig. 1(a,d) and Fig. 5) in seven BCDs implies they are Type-1 XUV disks. Note that all our observed FUV profiles reach 27.25 mag arcsec$^{-2}$ and below, beyond the optical radii. Whereas Type-2 XUVs do not necessarily have structured UV clumps but show recent and high star formation activity at large galactocentric radii. Type-2 XUV disks satisfy the following two criteria\cite{Thilkeretal2007}:
\medskip

\noindent (i) $FUV - K_{s} \le 4$ \\
\noindent (ii) ${A_{LSB}} \geq 7\times {A_{K80}}$ 

\noindent where FUV$-$K$_{s}$ color is computed in the LSB zone (defined as the area between 'threshold' SFRD contour and the contour enclosing 80$\%$ of the total flux in K$_{s}$ band). $A_{LSB}$ is the area of the LSB zone in FUV while $A_{K80}$ refers to the area enclosing $80\%$ of the total flux in $K_s$~band. We refer to the aforementioned XUV disk criteria as ``Thilker's criteria'' throughout the text. Note that to check whether our BCDs are Type-2, we consider the LSB zone's outer radius to be $R_{out,FUV}$. Since FUV and K$_{s}$ band PSFs differ by more than a factor of 3, we compute the color using the intrinsic FUV models matched to the PSF of K$_{s}$ band imaging ($\sim$0.4 arcsec)\cite{Fontanaetal2014}. We find that GS3 ($FUV - K_{s}$ = 1.25, ${A_{LSB}} = 16.9{A_{K80}}$), GS6 ($FUV - K_{s}$ = 0.56, ${A_{LSB}} = 7.2{A_{K80}}$) and eight others, i.e., a total of 10 BCDs, satisfy Type-2 XUV criteria. In short, 11 BCDs in our sample are either Type-1 or Type-2 or both based on Thilker's criteria.

\subsubsection{Measuring XUV light}
\label{sec:XUVdef}
In addition to the ratio of FUV scalelength to optical, we measure FUV light fraction in the XUV region denoted by $\Delta {L_x}$:

\begin{equation}
    \scalebox{1.4}{
    $\Delta {L_x} = \frac{L_{x,FUV} -L_{x,opt}}{L_{x,FUV}},
    \label{eq:delta_l}$}
\end{equation}

\noindent where $L_{x,FUV}$ is the intrinsic light within the FUV extent (R$_{out,FUV}$), and $L_{x,opt}$ is the light within the blurred optical extent (R$_{out,opt}$) (see Fig.~1). These are defined as

\begin{equation}
\scalebox{1.3}{
    $L_{x,FUV}=\int_{0}^{R_{out,FUV}} { I_{FUV}(R) 2 \pi R dR}$ }
\label{eq:lx,fuv}
\end{equation}

\begin{equation}
\scalebox{1.3}{
    $L_{x,opt}=\int_{0}^{R_{out,opt}} { I_{FUV}(R) 2 \pi R dR}$ }
\label{eq:lx,fuv}
\end{equation}

\noindent where $I_{FUV}(R)$ is the intrinsic light at a radius $R$ and given by

\begin{equation}
\scalebox{1.3}{
    $I_{FUV}(R)=I_{0,FUV}~\exp{(-R/R_{h,FUV}} $)}.
\label{eq:lx,fuv}
\end{equation}
Here, $I_{0,FUV}$ is the intrinsic peak intensity and $R_{h,FUV}$ is the intrinsic FUV scale-length obtained from the profile fit. According to the above relation, $\Delta {L_x} = 0$ when $R_{out,FUV} = R_{out,opt}$. 
Values of $\Delta L_x$ are in Table~5.

\subsection{UV color and stellar population ages}
\label{sec:SB99model}
We use STARBURST99\cite{leitherer99} stellar population models to estimate an average stellar population age based on the observed $FUV-NUV$ colors of our BCDs. For this, we make use of an instantaneous star-burst and a continuous star-formation history (CSFH) with a Salpeter IMF\cite{Salpeter1955} (1-100 M$_{\odot}$) and solar metallicity (Z$_{\odot}$) along with the Calzetti \cite{calzetti2000} extinction law. 
 We then obtain the model $FUV-NUV$ colors up to 1 Gyr by convolving the reddened (using the $E(B-V)$ values) model spectra with the UVIT/F154W and N242W filters. We find that, for an  instantaneous burst, the mean age of the stellar population in the XUV disk of the BCDs is $\sim26$ Myr. In case of a CSFH, the mean stellar population can be $\sim$230 Myr old (excluding GS10 and GS11 where it would be $\geq$1 Gyr old). We also find that, in the case of lower metallicity, Z = 0.4 Z$_{\odot}$, the XUV disks could be $\sim$42-324 Myr old. We tabulate the model ages for the XUV disks in Table 6.

\subsection{Clumps: detection and stellar mass estimates}

    \begin{table}[!h]
        \centering
        \begin{tabular}{c c c c}
            \hline
             \multicolumn{4}{c}{SExtractor parameters} \\
             \specialrule{0.75pt}{0pt}{0pt}
             \multicolumn{2}{c}{DETECT MINAREA} & \multicolumn{2}{c}{5} \\
             \multicolumn{2}{c}{DETECT THRESH} &\multicolumn{2}{c}{\textbf{1.5}, 2, 2.5} \\
             \multicolumn{2}{c}{FILTER NAME} &\multicolumn{2}{c}{\textbf{gauss 1.5 3x3.conv}, gauss 2.0 3x3.conv} \\
             \multicolumn{2}{c}{DEBLEND NTHRESH}& \multicolumn{2}{c}{64} \\
             \multicolumn{2}{c}{DEBLEND MINCONT} &\multicolumn{2}{c}{1$\times$10$^{-6}$} \\
             \multicolumn{2}{c}{CLEAN PARAM}& \multicolumn{2}{c}{10} \\
             \multicolumn{2}{c}{BACK TYPE} &\multicolumn{2}{c}{MANUAL (locally obtained background as described above)} \\
             \specialrule{0.75pt}{0pt}{0pt}
             &&&\\
             \multicolumn{4}{c}{Noise Chisel parameters}\\
             \specialrule{0.75pt}{0pt}{0pt}
             \multicolumn{2}{c}{Detection}  & \multicolumn{2}{c}{Segmentation}\\
             \specialrule{0.75pt}{0pt}{0pt}
              erode & 1 & tilesize& 40,40\\ 
              erodengb& 8 & minskyfrac& 0.1\\
              tilesize& 40 ,40 & kernel& 2 pixel fwhm\\
              interpnumngb& 4 & interpnumngb& 4\\
              minskyfrac& 0.1 & snquant& 0.4\\
              meanmedqdiff& 0.05 & gthresh& 1\\
              detgrowquant& 0.9 & snminarea& 4\\
              noerodequant& 0.9 & objbordersn& 1.0\\
              dthresh& 0.1 & &\\
              snminarea& 3 & &\\
              sigmaclip& 3,50 & &\\
              qthresh& 0.8 & &\\
              snthresh& 3.0 & &\\
             \specialrule{0.75pt}{0pt}{0pt}
        \end{tabular}
        \caption{\textbf{Clump detection parameters.} The configuration for \textit{SExtractor} and Noise Chisel used to detect the 'clumps'. A detection threshold of 1.5$\sigma$ and gauss 1.5 pixel fwhm filter (boldface) has been used in our final calculations.}
        \label{tab:detectparams}
    \end{table}

We employ a threefold approach to identify young star-forming structures in the BCDs. For this, we run \textit{SExtractor}\cite{Bertin_Arnouts1996} and Noise Chisel\cite{akhlagi2019} on the FUV cutouts of the BCDs, to detect and identify sources, with the configurations presented in Table 2. Our third approach is visually motivated where both tools succeed to  detect but fail to deblend these structural irregularities in the XUV region. We estimate the S/N of the detections using the segmentation maps in both the cases of \textit{SExtractor} (solid yellow contours in Fig.~1, Fig. 5) and Noise Chisel (marked by magenta dashed contours in Fig.~1, Fig. 5). In addition, for the \textit{SExtractor} clumps we estimate their S/N based on the Kron-like apertures (see Table 4). For the visually identified sources, we estimate the S/N within a fixed elliptical aperture (marked by red ellipses in Fig.~1, Fig. 5). For all the sources detected and identified, we consider only those detections having S/N$\geq$3 (or $>5\sigma$ above the background) for subsequent analysis. Of these, the ones in the XUV region without any HST counterpart are referred to as 'clumps' and considered to be part of the host BCD. We follow this definition of clumps throughout the paper.

Note that the photometry of faint objects will be influenced by the background. For example, in GS3, we obtain a local background of 28.77 mag arcsec$^{-2}$ using \textit{SExtractor}\cite{Bertin_Arnouts1996}, as compared to a higher value to 28.53 mag arcsec$^{-2}$ (obtained as described in section ~\ref{data_descrip}). This however, did not have any drastic effect on the stellar masses of the clumps; the mass of the fainter clump (C3B) only differed by $\sim$0.2 dex. Similarly, varying the \textit{SExtractor} parameters also brings changes in the clump masses. For example, changing the detection threshold from 1.5 to 2$\sigma$ while keeping the smoothing kernel intact, the stellar mass of the clump C3A changes by $\sim$0.2 dex. On the other hand, changing the smoothing kernel from gauss 1.5 to gauss 2.0 (at fixed detection threshold=1.5$\sigma$) has a lesser effect; e.g., clump mass of C3A changes by $\sim$0.1 dex. Beyond a detection threshold $>$2 in SExtractor, most of the FUV clumps go undetected. Therefore, we finally use parameters (Table 2 in Methods) motivated by those used in existing deep and large scale surveys\cite{galametz2013, hathi2012} , especially to pick faint sources. 

\subsubsection{Clump significance:}

\begin{table}[!hb]
    \centering
    \begin{tabular}{ c c c c c c }
    \specialrule{0.75pt}{0pt}{0pt}    \multicolumn{3}{c}{GS3} & \multicolumn{3}{c}{GS6} \\
    \specialrule{0.75pt}{0pt}{0pt}
        x & y & S/N & x & y & S/N\\
    \specialrule{0.5pt}{0pt}{0pt}
        -3.87	& 15.86 &	3.24 $^{*}$ & 7.81	& 11.11 & 3.3 $^{\#}$ \\
        -13.55 &	-12.35 &	2.27 & -17.33 &	8.64 & 3.2 $^{*}$ \\
        -4.19	& -16.88 & 2.60 & -11.14 &	6.64 & 3.0 $^{*}$ \\
        -10.45 &	9.46 & 2.30 & 12.57	& 1.91 & 2.4 \\
        11.17	&-12.80 &	2.34 & 9.75 & -3.00 & 2.3 \\
        -8.36	& 12.54 &	1.86 & 9.36	& -6.63 & 2.3 \\
        -14.10 & -7.29 & 2.25 & 11.48	& -10.30 & 2.4 \\
        &&& -9.17	& 0.11 & 2.8 \\
        &&& 8.80	& 8.26 & 2.1 \\
    \specialrule{0.75pt}{0pt}{0pt}    
    \end{tabular}
    \caption{\textbf{S/N of \textit{SExtractor} detected sources around GS3 and GS6.} The coordinate columns represent relative positions with respect to the galaxy's center. All the above tabulated sources are shown in Extended Fig. 7 c,e with outlines.  Those with * have S/N$\geq$3 with an HST counterpart and are marked with cyan boxes, and the one with \# also has S/N$\geq$3 but has no HST counterpart and is marked with an orange box. }
    \label{tab:gs3-gs6-features}
\end{table}

We detect a total of 12 clumps: 4 with SExtractor, 2 with Noise Chisel, and 6 clumps by visual identification in the XUV regions of the BCDs.
In order to measure the statistical significance and frequency of our FUV clumps, we perform the following experiment. We choose a large patch (341x341 pixels = 20220 arcsec$^{2}$) of the GOODS-South image in FUV and run \textit{SExtractor} (and Noise Chisel) with the same settings as used to find the clumps in our BCDs. This has resulted in a detection of 625 (157) sources. Of these, 13 (11) FUV sources have S/N$\geq$3 and without any HST counterpart. This translates to a clump density of 0.0006429 (0.0005440)~arcsec$^{-2}$ in the patch. {\it If these clumps are due to background fluctuations alone, we would expect $\sim$0.3 (0.2) similar clumps using SExtractor (and Noise Chisel) in the XUV regions, having a total area of $\sim 458.5$~arcsec$^2$ around all BCDs.} When we randomly place elliptical apertures (same as red ellipses, shown in Fig.~1d and Fig. 5 on the same patch, we find 0 out of 1604 apertures to have S/N$\geq$3 and without any HST counterpart. We then visually identify 435 FUV sources without HST counterpart in the same patch (once all SExtractor detected sources are removed) and estimate their S/N. We find that none qualify the S/N$\ge3$ threshold. 

The S/N of the SExtractor detected sources outside the XUV regions of GS3 and GS6 are presented in Table~3, and outlined in the Fig.~7(c, e). We find that all 6 sources surrounding GS3 and 6 of the sources (positive peaks) in GS6 have $S/N < 3$; these are all cyan circles in the figure. The cyan boxes (having HST counterparts) and orange box without an HST counterpart have $S/N > 3$.
In order to estimate and compare the fluxes corresponding to the positive and negative peaks just outside the XUV region, we use fixed aperture of size equal to the FWHM of the FUV PSF. Positive fluxes are estimated at the positions of the Sextractor detection while we use visual inspection for locating the negative peaks and the same aperture to estimate the negative fluxes (see Fig.~7 c, e). Fig.~7 (d,f) show  the histograms for these fluxes in case of GS3 and GS6. The ratios of the positive peak flux to negative peak flux i.e., the peak-to-peak flux ratios are 3.6 and 2.6 for GS3 and GS6 respectively. These ratios are slightly on the higher side, possibly because the noise distribution in far-UV (being dominated by low-photon statistics) is Poissonian rather than symmetric Gaussian.

\begin{figure}[!htb]
    \centering
    \includegraphics[width=\textwidth]{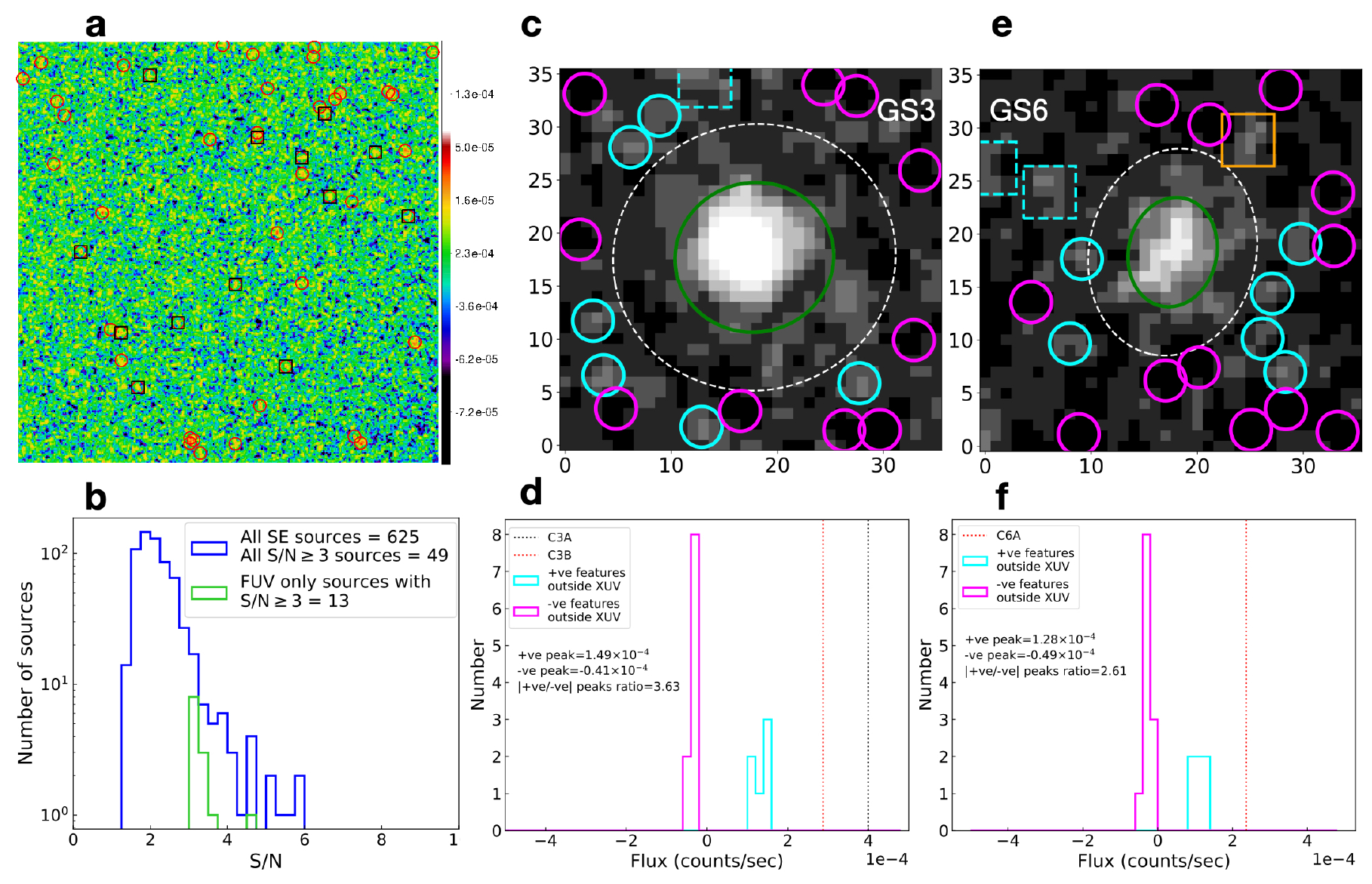}
    \caption{\textbf{Test for clump significance:} \textbf{a.} Patch of sky of size 341$\times$341 pixels devoid of bright and large sources in the GOODS-South field. Red circles denote sources having S/N$\geq$3 and their number is 49. Of these, 13 are marked with black circles that do not have any HST counterpart . \textbf{b.} S/N histogram of sources detected in the patch shown in \textbf{a}. Panel~\textbf{c, e} GS3 and GS6 with the same size as presented in Fig.~1. Cyan circles and boxes mark the SExtractor detected sources. Sources having S/N$<$3 are marked with cyan circles. Sources having S/N$\geq$3 and HST counterparts are marked with cyan dashed squares. The orange square has S/N$\geq3$ and no HST counterpart. Magenta circles mark noise dips below a flux value of zero. Panel~\textbf{d, f} represent flux (measured within a fixed aperture of size 0.7" radius) histograms corresponding to panel~\textbf{c, e.}}
    \label{fig:fig7}
\end{figure}

\subsubsection{Stellar mass estimates:}

\begin{table}[!h]
    \centering
    \begin{tabular}{c c c c c c c c c c }
    \hline
        & & & & & & & & &\\[-2ex]
        Clump ID & S/N & MAG & A$_{FUV}$ & SFRD$_{clump}$ & M$_{clump}$ & T$_{insprial}$ & ${\dot{M}}_{clump}$ & $\tau_{clump}$ & $\tau_{cc}$\\
          &  & mag & mag & M$_{\odot}$~yr$^{-1}$kpc$^{-2}$ & $\log$ M$_{\odot}$ & Gyr & M$_{\odot}$~Gyr$^{-1}$ & Gyr & Gyr\\
        \specialrule{0.75pt}{0pt}{0pt}
        $^\dagger$C1A & 3.3 & 26.18 & 1.21 & 7.63$\times$10$^{-4}$ & 6.42 &12.6 &0.21$\times$10$^{6}$ &12.5 &0.9 \\
   \specialrule{0.75pt}{0pt}{0pt}
    $^{\# \dagger}$C3A & 4.3 (4.8) &  25.30 
  & 2.00    &  4.98$\times$10$^{-4}$  & 6.84    & 5.6 & &  &1.0 \\
    $^{\#}$C3B &   3.7 (4.4) &  25.46  
  & 2.00    &  4.82$\times$10$^{-4}$ &  6.77  & 6.0 &2.3$\times$10$^{6}$ & 5.5 &1.9 \\
  \hline
    $^{*}$C6A &   3.2    & 26.84 
  & 0.86  & 1.16$\times$10$^{-3}$  &    6.49 & 7.5 & 0.43$\times$10$^{6}$ &7.2  &1.5 \\
  \hline
    $^{*}$C10A &   3.2    & 26.91 & 1.21 & 1.31$\times$10$^{-3}$ & 6.50 &  8.3 &0.40$\times$10$^{6}$ &7.9 &1.8 \\
  \hline
  $^{\# \dagger}$C12A    &  3.2 (3.6)     &  26.40
  & 1.30 & 5.44$\times$10$^{-4}$ & 5.90   &11.8 & & &1.1 \\
  $^{\dagger }$C12B    &   3.4    &   26.53 
        & 1.30 & 6.90$\times$10$^{-4}$  &  5.85   &54.8 & & &0.6 \\
  $^{*}$C12C     &   4.0    &  26.28
  & 1.30 & 9.42$\times$10$^{-4}$ & 5.95   & 26.5&  & &0.5 \\
  $^{*}$C12D    &  3.2     &   26.67 
        & 1.30 & 1.30$\times$10$^{-4}$  &  5.80  & 72.0 &0.13$\times$10$^{6}$ &23.2 &0.4 \\
\hline
  $^{*}$C13A    &   3.0    &  27.04
  & 1.63 & 2.93$\times$10$^{-3}$ & 6.76   & 18.6 &  &  &0.9 \\
  $^{*}$C13B  &   3.4    &   26.64 
        & 1.63 & 7.89$\times$10$^{-3}$  &  6.92   & 13.1 &0.96$\times$10$^{6}$ &14.6 &0.7 \\
\hline
  $^{\# \dagger}$C14A    &   3.8 (3.0)    &  26.26 
  & 2.95    & 9.49$\times$10$^{-4}$  & 6.83 & 2.4 &3.0$\times$10$^{6}$ &2.2 &1.8 \\
      \hline
    \end{tabular}
    \caption{\textbf{Full FUV-clump analysis in our sample of BCDs.} \textbf{Column 1:} Clump IDs identified by \textit{SExtractor} ($\#$), Noise Chisel ($\dagger$) and visual inspection ($*$). \textbf{Column 2:} Measured S/N of the clumps (in \textit{SExtractor} Kron-like apertures). \textbf{Column 3:} Clump magnitudes using \textit{SExtractor} within Kron-like apertures (using detected pixel flux in the case of Noise Chisel and aperture fluxes in the case of visually identified clumps). \textbf{Column 4:}  Internal dust extinction estimated using the UV-slope ($\beta$). \textbf{Column 5:} Clump star formation rate surface density. \textbf{Column 6:} Clump stellar mass. \textbf{Column 7:} Inspiral time of the clumps. \textbf{Column 8:} Net clump accretion rate. \textbf{Column 9:} Total clump accretion timescale. \textbf{Column 10:} Clump-clump collision timescale.}
    \label{tab:clump_mags}
\end{table}

 We convert the measured FUV fluxes of the clumps into SFRs\cite{kennicutt12} and estimate their stellar masses (see Table~4) assuming a constant SFR for 100 Myrs as:
 \begin{equation}
    SFR~(M_{\odot}~yr^{-1}) = 1.4~\times~10^{-28}~L_{\nu}~(ergs~s^{-1}~Hz^{-1})
 \end{equation}

Note that all the FUV-flux-derived stellar masses, quoted throughout the paper, have been corrected for internal dust extinction with the help of the UV-slope ($\beta$), obtained by fitting a straight line within 1268\AA\ - 2580\AA\  to the intrinsic Spectral Energy Distribution (SED) models. The color excess is then obtained as\cite{Reddyetal2018}
 
 \begin{equation}
     E(B-V) = \frac{1}{4.684} [\beta + 2.616]
 \end{equation}

Due to the sensitivity of $\beta$ to metallicity (Z), stellar ages and the SFH\cite{Zeimannetal2015,Bouwens2016a}, the derived extinction can be biased. We use intrinsic SED models with Z = Z$_{\odot}$ and 0.4~Z$_{\odot}$, and find $\Delta\beta$ up to 0.26 or equivalently, $\Delta E(B-V)$ up to 0.06 mag only. We proceed with $E(B-V)$ at Z$_{\odot}$ and use STARBURST99 models (section \ref{sec:SB99model}) to model clump masses. For a 100 Myr old instantaneous burst, our mean clump mass is in good agreement with the mean model clump mass. Model masses gradually deviate by $\sim0.5$ dex to 3.5 dex for a 200 Myr to 1 Gyr old burst. A difference of $\sim$0.9 dex is seen for a 100 Myr old CSFH.

\subsection{In-spiral timescale}
\label{sec:inspiral}
In order to estimate the inspiral time scale due to the dynamical friction from the dark matter halo, we assume Binney's logarithmic spherical potential \cite{binney08} given by:

\begin{equation}
    \Phi_{L}(r) = \frac{V_{0}^2}{2} \ln[{R_{c}^2 +r^2}], 
\end{equation}

\noindent where $R_c$ is the core radius of the dark matter halo and $V_{0}$ is the asymptotic velocity. We then integrate the equation $d\vec{L_{z}}/dt = \vec{r} \times \vec{F}_{df}$, where $\vec{F}_{df}$ is the dynamical friction force (see Eq.~\ref{eq:dvdt}) due to the halo, to obtain the in-spiral timescale. The Coulomb factor appearing in Eq.~\ref{eq:dvdt} can be written as

 \begin{equation}
      \ln \Lambda = \ln \left(\frac{R_{gal}}{R_{min}}\right) \sim \ln\left(\frac{M_{dyn}}{M_{c}}\right) \sim \ln \left(\alpha \frac{M_{*}}{M_{c}} \right),
  \end{equation}
  
\noindent where $R_{\rm gal}$ is the size of the galaxy and $R_{\rm min}\simeq GM_{\rm c}/V^2$ is the strong encounter radius. Then the in-spiral timescale is given by

\begin{equation}
    \frac{T_{insprial}}{\tau_{orb}}= {\frac{{{\alpha {M_{*}}/{M_{c}}} }}{\ln (\alpha M_{*}/M_{c})}} \frac{X_{dm}^2}{\xi(X_{dm})}  \frac{I_{dm}(p,R_{out},R_{in})}{2\pi ({R_{out}/{p R_d}})^2},
    \label{eq:Tinspiral}
\end{equation}

\noindent where,
\begin{equation}
    I_{dm}(p,R_{out},R_{in}) = \int_{\frac{R_{out}}{p R_d}}^{\frac{R_{in}}{p R_d}} {\frac{2+x^2}{3+x^2} \frac{x^2}{\sqrt{1+x^2}}} dx.
    \label{eq:dmintegral}
\end{equation}

\noindent The indefinite integral $I_{dm}$ (above Eq.~\ref{eq:dmintegral}) can be written as $I_{dm} = I_{dm,1} + 3 I_{dm,2}$ such that, 

\begin{equation}
    I_{dm,1} = \frac{1}{2}\left[x\sqrt{1+x^2} - 3 \ln(x+ \sqrt{1+x^2})\right],
\end{equation}

\noindent and

\begin{equation}
    I_{dm,2} = \frac{1}{2\sqrt{6}}\left[\ln(|\frac{2 x}{\sqrt{1+x^2}} + \sqrt{6}|) - \ln(|\frac{2 x}{\sqrt{1+x^2}} - \sqrt{6}|)\right]
\end{equation}

\noindent In the above equations, $R_{d}$ is the disk scale-length; $X_{dm}$ and $\tau_{orb}$ (orbital timescale) are given by, 

\begin{equation}
    X_{dm}^2 \equiv \frac{V_{c0}^2}{V_{0}^2}={\frac{\alpha}{\alpha -1} h(R_{out}/{p R_{d}})}, \: \: \: and \: \: \: \tau_{orb} = \frac{2 \pi R_{out}^{3/2}}{\sqrt{G \alpha M_{*}}}
\end{equation}

\noindent where $V_{c0}$ is the circular velocity calculated at the location of the clump ($R_{out}$) due to the total dynamical mass ($M_{dyn} =\alpha M_{*}$) of the galaxy, with $\alpha=2.7$ for all BCDs. This, in turn, fixes the dark matter halo mass as $M_{dark} =(\alpha -1) M_{*}$;  and $R_c = p R_d$. In all our calculations, we consider $p=2$ case. Increasing the parameter $p$ reduces the in-spiral timescale. Varying the parameter $p$ either to $1$ or $3$ changes the in-spiral timescale by $\sim 10\%$.
The function $h(y)=y^2/(1 +y^2)$ represent the radial dependence of the circular velocity curve for the assumed dark matter halo potential. For GS3 with $M_{*} = 3.78 \times 10^8 M_{\odot}$, $\tau_{orb} = 1.7$~Gyr at $R_{out} = 7$~kpc.

\begin{table}[!hb]
\small{
\centering
\caption{\textbf{Various metrics derived for the XUV disks of the BCDs.} \textbf{Column 2:} Total stellar mass of the galaxy. \textbf{Column 3:} Young stellar mass of the galaxy. \textbf{Column 4, 5:} FUV-K$_{s}$ color and fraction of LSB area measured to check for Type 2 XUV disk. \textbf{Column 6:} Type of XUV disk as per Thilker's criteria, Type-1 (T1) and Type-2 (T2). \textbf{Column 7:} FUV fraction in XUV region. \textbf{Column 8:} Observed FUV-NUV color in the XUV region of the BCDs. \textbf{Column 9:} Intrinsic, profile-integrated SFRD values in the XUV region. \textbf{Column 10:} XUV disk evolution timescale. \textbf{Column 11:} Galaxy star formation time. The quoted uncertainties are 1$\sigma$.}
\label{tab:XUV_detection}
\begin{tabular}{c c c c c c c c c c c}
\hline	

&&&&&&&&&&\\[-2.0ex]

BCD   & M$_{*,total}$ & M$_{*,young}$  &FUV-K$_{s}$ & $\frac{A_{LSB}}{A_{K80}}$ & XUV &$\Delta L_{x}$ & {FUV-NUV} & ${SFRD}_{XUV}$ & $\tau_{XUV}$ & $\tau_{SF}$ \\
\specialrule{0.75pt}{0pt}{0pt}
& log M$_{\odot}$ & log M$_{\odot}$ & mag & & Type &\% & mag & M$_{\odot}$yr$^{-1}$kpc$^{-2}$ &  Gyr & Gyr \\
\specialrule{0.75pt}{0pt}{0pt}														
GS1	&	8.13	&	7.4		& 0.08  & 17.6 &	T1,T2&12.3$\pm$5.7& -1.78$\pm$0.29	&	1.3$\pm0.49$$~\times$10$^{-4}$	&	14.6 & 1.7 \\

GS2	&	8.73	&	8.12	& -0.03  &  13.4 &T2&	34.5$\pm$5.6	&  0.00$\pm$0.18 &	2.4$\pm0.49$$~\times$10$^{-4}$	&	--  & 1.8 \\		
														
GS3	&	8.57	&	7.83	& 1.25  & 16.9 &T1,T2	&3.2$\pm$0.1	& -0.37$\pm$0.21	&1.8$\pm0.43$$~\times$10$^{-5}$	&	6.6	& 3.4\\		
														
GS4	&	9.02	&	7.9	& 1.53  & 15.0 &T2	&11.6$\pm$3.5	& -0.42$\pm$0.33	&8.7$\pm2.30$$~\times$10$^{-5}$	&	-- & 8.1	\\		
														
GS5	&	8.92	&	7.87	& 3.24  & 6.5 &--	&0.1$\pm$0.1	& -0.92$\pm$1.45	&1.1$\pm8.58$$~\times$10$^{-5}$	&	--	& 11.9\\		
														
GS6	&	8.65	&	7.6		&  0.56 & 7.2 &T1,T2	&33.6$\pm$6.7	& -0.44$\pm$0.24	&2.3$\pm0.48$$~\times$10$^{-4}$	&	31.3 & 2.5\\		
														
GS7	&	8.70	&	8.43	& -0.87  & 34.1 &T2	&1.8$\pm$1.1	& -0.29$\pm$0.22	&1.8$\pm0.83$$~\times$10$^{-5}$	&	--  & 1.1  \\		
														
GS10	&	8.48	&	7.74&  -0.41 & 10.3 	&	T1,T2&39.2$\pm$5.4	&	0.11$\pm$0.20 &3.9$\pm0.64$$~\times$10$^{-4}$	&	54.9 & 1.7   \\		
														
GS11	&	8.99	&	8.69	& -0.01  & 22.2 & T2	&3.0$\pm$0.1	& 0.14$\pm$0.13	&6.4$\pm1.10$$~\times$10$^{-5}$	&	--  & 0.8  \\		
														
GS12	&	8.66	&	7.57	& 3.68  & 6.1 &T1	&0.1$\pm$0.0	& -0.56$\pm$0.20	&5.8$\pm0.13$$~\times$10$^{-6}$	&	26.3 & 4.2	\\		
														
GS13	&	8.59	&	8.07	& -0.61  & 20.1 &T1,T2	&21.9$\pm$6.4	& -1.15$\pm$0.28	&2.7$\pm0.74$$~\times$10$^{-4}$	&	26.6 & 1.6	\\		
														
GS14	&	8.73	&	8.02	& 2.17  & 14.0 &	T1,T2&4.2$\pm$1.4	& 0.20$\pm$0.29	&2.4$\pm0.63$$~\times$10$^{-5}$	&	3.7 & 7.7\\		

\specialrule{0.75pt}{0pt}{0pt}
\end{tabular}
}
\end{table}

\subsubsection{Clump accretion rate and timescale:}
\noindent Based on the clump mass conservation, we calculate the net clump accretion rate in the galaxy as the sum of individual clump accretion rates due to the dynamical friction alone as

\begin{equation}
    {\dot{M}}_{clump} = \sum_{j=1}^{n}{\frac{M_{c,j}}{T_{inspiral,j}}},
\end{equation}

\noindent where n denotes the total number of detected clumps in the galaxy; $T_{inspiral,j}$ denotes inspiral time for $j^{th}$ clump.
In GS3, ${\dot{M}}_{clump} = 2.3 \times 10^6$~M$_{\odot}$~Gyr$^{-1}$. We utilize the net clump accretion rate to estimate the timescale for the clumps to reach the optical disk of the galaxy as:

\begin{equation}
 \tau_{clump} = {M_{clump}}/{{\dot{M}}_{clump}}
 \end{equation}

In addition, we estimate the timescale required for the clumps to transfer an amount of mass equal to the young stellar mass associated with the XUV disk into the optical region of the BCDs. We first note that galaxies with low $\Delta L_x$, in the range of a few percent, have XUV disk masses that are less than the summed clump masses. The average radial profile could be flatter in the outer regions where the clumps are, making it a Type III profile. To account for this in the timescale for outer disk evolution, we add the summed clump masses to the extrapolated XUV masses that come from the intrinsic fits in those 6 cases where $\Delta L_x$ is low. We do not add the clump masses to the other cases. This difference in the two cases is evident in the FUV images also because the low-$\Delta L_x$ cases have outer clumps with much less contrast to the rest of the outer disk than the high-$\Delta L_x$ cases. Thus we evaluate the XUV evolution timescales from the following equations and show them in Table~5:

\begin{equation}
    \tau_{XUV} = \frac{\Delta L_{x} M_{young}}{{\dot{M}}_{clump}} \;\;;\;\;(\Delta L_x>10\%)
\end{equation}

\begin{equation}
    \tau_{XUV} = \frac{\Delta L_{x} M_{young}+M_{clump}}{{\dot{M}}_{clump}}
    \;\;;\;\;(\Delta L_x<10\%)
\end{equation}

\subsubsection{Clump-clump collision timescale}
Clumps formed in the XUV disk will interact gravitationally and might merge to grow bigger, in which case they will fall faster to the central region of the host BCD\cite{Papaderosetal2002}. The clump infall time would essentially be determined by the clump-clump collision timescale. The collision cross section of a clump (assuming spherical shape) is simply $\sigma_{clump} = \pi R_{clump}^2$, where $R_{clump}$ is the radius of the clump. Then the mean free path of the clump is given by

\begin{equation}
    l_{clump} = \frac{1}{n_{clump} \sigma_{clump}}, 
\end{equation}

\noindent where $n_{clump}$ denotes the number density of the clumps within the XUV region. The clump-clump collision timescale, denoted as $\tau_{cc}$, relative to the orbit time can then be written as 
\begin{equation}
    \frac{\tau_{cc}}{\tau_{orb}} = \frac{1}{2} \frac{M_{clump}}{\sum{M_{clump}}}\frac{H}{R_{out}}\left(\frac{R_{XUV}}{R_{clump}}\right)^2  
\end{equation}

In the above equation, $R_{XUV}$ denotes the size of the XUV region; H is the thickness of the disk and the rest of the parameters have their usual meaning. In all our calculations, we assume $H/R_{XUV} = 0.2$, for simplicity. In both GS3 and GS6, the mean clump collision timescale is $\sim$1.5~Gyr ($\sim \tau_{orb}$). In other words, clumps in most BCDs will collide with another within an orbital timescale.
 
\begin{table}[!t]
\centering
\caption{\textbf{XUV ages based on observed $FUV-NUV$ color and stellar population synthesis.} \textbf{Column 2, 3:} Age estimates using an instantaneous burst and CSFH with Z=Z$_{\odot}$. \textbf{Column 4, 5:} Age estimates using a SF burst and continuous SF with 0.4 Z$_{\odot}$. We ignore GS1 due to extreme blue color and GS5 due to being a marginal case.}
\label{tab:ssp_ages}
\begin{tabular}{c c c c c }
\specialrule{0.75pt}{0pt}{0pt}	
\multirow{3}{*}{BCD} & \multicolumn{4}{c}{Ages (Myr)}\\
\specialrule{0.75pt}{0pt}{0pt}
& SF burst  & continuous SF & SF burst  & continuous SF \\
 & Z$_{\odot}$  & Z$_{\odot}$ & 0.4 Z$_{\odot}$  & 0.4 Z$_{\odot}$ \\
\specialrule{0.75pt}{0pt}{0pt}
GS1 & --    & --    &-- & -- \\
GS2 & 52.2  & 748.2 &  72.0 & 769.8\\
GS3 & 7.2   & 106.2 & 19.1  & 196.8\\
GS4 & 6.7   & 73.6  & 17.4  & 155.3\\
GS5 &   --  & --    & -- & --\\
GS6 & 12.3  & 306.3 & 34.3  & 472.4\\
GS7 &   8.1 & 175.2 & 24.3  & 321\\
GS10    &   81.9    & $>$1000 & 98.2   & $>$1000\\
GS11    &   72.6    &   $>$1000    &   93.6    &   $\sim$1000\\
GS12    &   7.0 &   99.5    &   18.8    &   190.0 \\
GS13    &   1.8 &   2.6 &   3.0 &   4.7 \\
GS14    &   12.3    &   327.5   &   37.5    &   482.1   \\
\specialrule{0.75pt}{0pt}{0pt}
\end{tabular}
\end{table}

\bigskip
\noindent {\bf Data availability}\\
The HST imaging data are available at {\url{https://archive.stsci.edu/hlsps/hlf/v1.5/}} and 3D-HST catalog is available at {\url{https://3dhst.research.yale.edu/Data.php}}. The HAWK-I K$_{s}$ band data are available at ESO Science Archive Facility (\url{http://archive.eso.org/scienceportal/home}). The original level 1 far-UV data observed by UVIT/AstroSat is available for download from the ISSDC site at \url{https://astrobrowse.issdc.gov.in/astro\_archive/archive/Home.jsp}.

\noindent {\bf Code availability}\\
We have used standard data reduction tools in Python, IRAF, and the publicly available code SExtractor ({\url{https://www.astromatic.net/software/sextractor}}) and PROFILER ({\url{https://github.com/BogdanCiambur/PROFILER/}}) for this study. We also have used the  MPFIT routine translated to Python language here ({\url{https://people.ast.cam.ac.uk/~rcooke/python/packages/mpfit.py}}). The pipeline used to process the Level 1 AstroSat/UVIT data can be downloaded from {\url{http://astrosat-ssc.iucaa.in}}.


\begin{thebibliography}{1}

\bibitem{GildePazetal2003}
{{Gil de Paz}, A., {Madore}, B.~F. \& {Pevunova}, O.} Palomar/Las Campanas Imaging Atlas of Blue Compact Dwarf Galaxies. I. Images and Integrated Photometry. \textit{\apjs}, \textbf{147}, 29-59 (2003)

\bibitem{Kunth-Ostlin2000}
{{Kunth}, D. \& {{\"O}stlin}, G.} The most metal-poor galaxies. \textit{\aapr}, \textbf{10}, 1-79 (2000)

\bibitem{hunter06}
{{Hunter}, D.~A., \& {Elmegreen}, B.~G.} Broadband Imaging of a Large Sample of Irregular Galaxies. {\it ApJS}, {\bf 162}, 49-79 (2006)

\bibitem{Ostlinetal2021}
{{{\"O}stlin}, G., {Rivera-Thorsen}, T. E., and {Menacho}, V.}\textit{et al.} The Source of Leaking Ionizing Photons from Haro11: Clues from HST/COS Spectroscopy of Knots A, B, and C. \textit{\apj}, \textbf{912}, 155 (2021)

\bibitem{Calvietal2014}
{{Calvi}, V., {Stiavelli}, M., {Bradley}, L., {Pizzella}, A. \& {Kim}, S.} The Effect of Surface Brightness Dimming in the Selection of High-z Galaxies. \textit{\apj}, \textbf{796},102, (2014)

\bibitem{Tandonetal2017a}
{{Tandon}, S.~N.} \textit{et al.} In-orbit Performance of UVIT and First Results. \textit{\JApA}, \textbf{38}, 28 (2017)

\bibitem{Singhetal2014}
{{Singh}, K.~P.} \textit{et al.} ASTROSAT mission. \textit{\procspie}, {\bf 9144}, 91441S (2014)

\bibitem{Hunteretal2011}
{{Hunter}, D.~A.} \textit{et al.} The Outer Disks of Dwarf Irregular Galaxies. \textit{\aj}, \textbf{142}, 121 (2011)

\bibitem{Lianetal2015}
{{Lian}, J.~H., {Kong}, X., {Jiang}, N., {Yan}, W. \& {Gao}, Y.~L.} Surface brightness profiles of blue compact dwarf galaxies in the GOODS-N and GOODS-S field. \textit{\mnras}, \textbf{451}, 1130-1140 (2015)

\bibitem{galametz2013}
{{Galametz}, A.} \textit{et al.} CANDELS Multiwavelength Catalogs: Source Identification and Photometry in the CANDELS UKIDSS Ultra-deep Survey Field. \textit{\apjs}, \textbf{206}, 10 (2013)

\bibitem{Schlafly-Finkbeiner2011}
{{Schlafly}, E.~F. \& {Finkbeiner}, D.~P.} Measuring Reddening with Sloan Digital Sky Survey Stellar Spectra and Recalibrating SFD. \textit{\apj}, \textbf{737}, 103 (2011)

\bibitem{boyleetal1998}
{{Boyle}, B.~J., {Shanks}, T. \& {Peterson}, B.~A.} The evolution of optically selected QSOs - II. \textit{\mnras}, \textbf{235}, 935-948 (1988)

\bibitem{Momchevaetal2016}
{{Momcheva}, I.~G.} \textit{et al.} The 3D-HST Survey: Hubble Space Telescope WFC3/G141 Grism Spectra, Redshifts, and Emission Line Measurements for \raisebox{-0.5ex}\textasciitilde 100,000 Galaxies, \textit{\apjs}, \textbf{225}, 27 (2016)

\bibitem{Thilkeretal2007}
{{Thilker}, D.~A.} \textit{et al.} A Search for Extended Ultraviolet Disk (XUV-Disk) Galaxies in the Local Universe. \textit{\apjs}, \textbf{173},  538 (2007)

\bibitem{Herrmannetal2013}
{{Herrmann}, K.~A., {Hunter}, D.~A. \& {Elmegreen}, B.~G.} Surface Brightness Profiles of Dwarf Galaxies. I. Profiles and Statistics. \textit{\aj}, \textbf{146}, 104 (2013)
    
\bibitem{kennicutt12} 
{{Kennicutt}, R.C., \& {Evans}, N.J.} Star Formation in the Milky Way and Nearby Galaxies. \textit{\araa}, \textbf{50}, 531-608 (2012)

\bibitem{hunter10}
{{Hunter}, D.~A., {Elmegreen}, B.~G. \& {Ludka}, B.~C.} Galex Ultraviolet Imaging of Dwarf Galaxies and Star Formation Rates. {\it Astron. J.}, {\bf 139}, 447-475 (2010)

\bibitem{jachym2014}
{{J{\'a}chym}, P., {Combes}, F., {Cortese}, L., {Sun}, M. \& {Kenney}, J.~D.~P.} Abundant Molecular Gas and Inefficient Star Formation in Intracluster Regions: Ram Pressure Stripped Tail of the Norma Galaxy ESO137-001. \textit{\apj}, \textbf{792}, 11 (2014)

\bibitem{Zavadsky2014}
{{Dessauges-Zavadsky}, M., {Verdugo}, C., {Combes}, F. \& {Pfenniger}, D.} CO map and steep Kennicutt-Schmidt relation in the extended UV disk of M 63. \textit{\aap}, \textbf{566}, A147 (2014)

\bibitem{Bicalhoetal2019}
{{Bicalho}, I.~C., {Combes}, F., {Rubio}, M., {Verdugo}, C. \& {Salome}, P.} ALMA CO(2-1) observations in the XUV disk of M83. \textit{\aap}, \textbf{623}, A66 (2019)

\bibitem{hunter12}
{{Hunter}, D.~A.} \textit{et al.} Little Things. \textit{\aj}, {\bf 144}, 134-163 (2012)  

\bibitem{boissieretal2012}
{{Boissier}, S.} \textit{et al.} The GALEX Ultraviolet Virgo Cluster Survey (GUViCS). II. Constraints on star formation in ram-pressure stripped gas. \textit{\aap}, \textbf{545}, A142 (2012)

\bibitem{leitherer99}
{{Leitherer}, C.} \textit{et al.} Starburst99: Synthesis Models for Galaxies with Active Star Formation. \textit{\apjs}, \textbf{123}, 3-40 (1999)

\bibitem{elmegreen2009}
{{Elmegreen}, D.~M.} \textit{et al.} Clumpy Galaxies in Goods and Gems: Massive Analogs of Local Dwarf Irregulars. \textit{\apj}, \textbf{701}, 306-329 (2009)

\bibitem{bell01}
{{Bell}, E. \& {de Jong}, R. S.} Stellar Mass-to-Light Ratios and the Tully-Fisher Relation, \textit{\apj}, \textbf{550}, 212-229, (2001)

\bibitem{Elmegreenetal2012}
{{Elmegreen}, B.~G., {Zhang}, H.-X., \& {Hunter}, D.~A.} In-spiraling Clumps in Blue Compact Dwarf Galaxies. \textit{\apj}, \textbf{747}, 105-114 (2012)

\bibitem{binney08}
{{Binney}, J., \& {Tremaine}, S.} {\it Galactic Dynamics}, Princeton University Press (2008)   

\bibitem{Ceverinoetal2010}
{{Ceverino}, D., {Dekel}, A. \& {Bournaud}, F.} High-redshift clumpy discs and bulges in cosmological simulations. \textit{\mnras}, \textbf{404}, 2151-2169 (2010)

\bibitem{Zhangetal2012}
 {{Zhang}, H-X., {Hunter}, D.~A., {Elmegreen}, B.~G., {Gao}, Y. \& {Schruba}, A.} Outside-in Shrinking of the Star-forming Disk of Dwarf Irregular Galaxies. \textit{\aj}, \textbf{143}, 47 (2012)
 
\bibitem{Schulte-Ladbecketal1999}
 {{Schulte-Ladbeck}, R.~E., {Hopp}, U., {Crone}, M.~M. \& {Greggio}, L.} A Stellar Population Gradient in VII ZW 403: Implications for the Formation of Blue Compact Dwarf Galaxies. \textit{\apj}, \textbf{525}, 709-719 (1999)


\end{thebibliography}

\bigskip
\medskip
\noindent{\bf References:}
\begin{enumerate}\addtocounter{enumi}{32}



\bibitem{oke1974}
{{Oke}, J.~B.} Absolute Spectral Energy Distributions for White Dwarfs. \textit{\apjs}, \textbf{27}, 21 (1974)

\bibitem{Skeltonetal2014}
{{Skelton}, R. E.} \textit{et al.} 3D-HST WFC3-selected Photometric Catalogs in the Five CANDELS/ 3D-HST Fields: Photometry, Photometric Redshifts, and Stellar Masses. \textit{\apjs}, \textbf{214}, {24} (2014)

\bibitem{Illingworthetal2016}
{{Illingworth}, G.} \textit{et al.} The Hubble Legacy Fields (HLF-GOODS-S) v1.5 Data Products: Combining 2442 Orbits of GOODS-S/CDF-S Region ACS and WFC3/IR Images. \textit{ArXiv e-prints}, \textbf{1606.00841}, (2016)

\bibitem{Fontanaetal2014}
{{Fontana}, A.} \textit{et al.} The Hawk-I UDS and GOODS Survey (HUGS): Survey design and deep K-band number counts. \textit{\aap}, \textbf{570}, A11 (2014)

\bibitem{Sahaetal2020}
{{Saha}, K.} \textit{et al.} AstroSat detection of Lyman continuum emission from a z = 1.42 galaxy. \textit{\natast}, \textbf{1185-1194}, 4 (2020)

\bibitem{Xuetal2005}
{{Xu}, C. Kevin} \textit{et al.} Number Counts of GALEX Sources in Far-Ultraviolet (1530 {\r{A}}) and Near-Ultraviolet (2310 {\r{A}}) Bands. \textit{\apjl}, \textbf{619}, L11-L14 (2005)

\bibitem{Bertin_Arnouts1996}
{{Bertin}, E. \& {Arnouts}, S.} SExtractor: Software for source extraction. \textit{\aaps}, \textbf{117}, 393-404 (1996)

\bibitem{Schlegeletal1998}
{{Schlegel}, D.~J., {Finkbeiner}, D.~P. \& {Davis}, M.} Maps of Dust Infrared Emission for Use in Estimation of Reddening and Cosmic Microwave Background Radiation Foregrounds. \textit{\apj}, \textbf{500}, 525-553 (1998)

\bibitem{Bianchi2011}
{{Bianchi}, L.} GALEX and star formation. \textit{\apss}, \textbf{335}, 51-60 (2011)

\bibitem{Tody1986}
{{Tody}, D.} The IRAF Data Reduction and Analysis System. \textit{\procspie}, \textbf{627}, 733 (1986)

\bibitem{Jedrzejewski1987}
{{Jedrzejewski}, R.~I.} CCD surface photometry of elliptical galaxies. I - Observations, reduction and results. \textit{\mnras}, \textbf{226}, 747-768 (1987)

\bibitem{Ciambur2016}
{{Ciambur}, B. C.} Profiler - A Fast and Versatile New Program for Decomposing Galaxy Light Profiles. \textit{\pasa}, \textbf{33}, e062 (2016)

\bibitem{Erwinetal2008}
{{Erwin}, P., {Pohlen}, M. \& {Beckman}, J.~E.} The Outer Disks of Early-Type Galaxies. I. Surface-Brightness Profiles of Barred Galaxies. \textit{\aj}, \textbf{135}, 20-54 (2008)

\bibitem{MPFIT}
{{Markwardt}, C.~B.} Non-linear Least-squares Fitting in IDL with MPFIT. \textit{Astronomical Data Analysis Software and Systems XVIII ASP Conference Series}, \textbf{411}, 251 (2009)

\bibitem{More1981}
{{Mor{\'e}}, J.~J., {Garbow}, B.~S. \& {Hillstrom}, K.~E.} Testing Unconstrained Optimization Software. \textit{ACM Transactions on Mathematical Software} \textbf{7(1)}, 17-41 (1981)

\bibitem{Salpeter1955}
{{Salpeter}, E.~E.} The Luminosity Function and Stellar Evolution. \textit{\apj}, \textbf{121}, 161 (1955)

 \bibitem{calzetti2000}
 {{Calzetti}, D.} \textit{et al.} The Dust Content and Opacity of Actively Star-forming Galaxies. \textit{\apj}, \textbf{533}, 682-695 (2000)

 \bibitem{akhlagi2019}
 {{Akhlaghi}, M.} Carving out the low surface brightness universe with NoiseChisel. \textit{arXiv e-prints}, \textbf{arXiv:1909.11230} (2019)

\bibitem{hathi2012}
{{Hathi}, N. P., {Mobasher}, B., {Capak}, P., {Wang}, W.-H. \& {Ferguson}, H. C.} Near-infrared Survey of the GOODS-North Field: Search for Luminous Galaxy Candidates at z$>\sim$6.5.
\textit{\apj}, \textbf{757}, 43 (2012)

\bibitem{Reddyetal2018}
{{Reddy}, N.~A.} \textit{et al.} The HDUV Survey: A Revised Assessment of the Relationship between UV Slope and Dust Attenuation for High-redshift Galaxies. \textit{\apj}, \textbf{853}, 56 (2018)

\bibitem{Zeimannetal2015}
{{Zeimann}, G.~R.} \textit{et al.} The Dust Attenuation Curve versus Stellar Mass for Emission Line Galaxies at z \raisebox{-0.5ex}\textasciitilde 2. \textit{\apj}, \textbf{814}, 162 (2015)

\bibitem{Bouwens2016a}
{{Bouwens}, R.~J.} \textit{et al.} ALMA Spectroscopic Survey in the Hubble Ultra Deep Field: The Infrared Excess of UV-Selected z = 2-10 Galaxies as a Function of UV-Continuum Slope and Stellar Mass. \textit{\apj}, \textbf{833}, 72 (2016)

\bibitem{Papaderosetal2002}
 {{Papaderos}, P.} \textit{et al.} The blue compact dwarf galaxy I Zw 18: A comparative study of its low-surface-brightness component. \textit{\aap}, \textbf{393}, 461-483 (2002) 

\end{enumerate}
\end{document}